%% file: main_new.tex
\documentclass[a4paper,11pt]{article}
\usepackage{jheppub}
\usepackage{lineno}
\usepackage{gensymb}
\usepackage{slashed}
\pdfoutput=1
\usepackage{color,graphicx,slashed,multirow,hyperref,amssymb,amsmath,cleveref}
\usepackage{braket}
\usepackage[utf8]{inputenc}
\usepackage[section]{placeins}
\usepackage{xcolor,colortbl}
\usepackage{subcaption}
\usepackage[shortlabels]{enumitem}
\usepackage{soul}
\AtBeginDocument{\usepackage{booktabs}}
\renewcommand{\arraystretch}{1.2}

\newcommand{\keV}{{\rm\ keV}}
\newcommand{\MeV}{{\rm\ MeV}}
\newcommand{\GeV}{{\rm\ GeV}}

\author[a]{Nakorn Thongyoi,}
\author[a]{Chakrit Pongkitivanichkul}
\affiliation[a]{Khon Kaen Particle Physics and Cosmology Theory Group (KKPaCT),\\
  Department of Physics, Faculty of Science, Khon Kaen University,\\
  123 Mitraphap Rd, Khon Kaen 40002, Thailand}
\emailAdd{nakoth@kku.ac.th}
\emailAdd{chakpo@kku.ac.th}

\input{abstract}

\begin{document}
\maketitle
\flushbottom

\input{intro}
\input{sec_model}
\input{sec_regimes}
\input{sec_consistency}
\input{sec_cannibal}
\input{sec_relocated}
\input{sec_tied}
\input{sec_pheno}
\input{sec_conclusions}

\acknowledgments
This research has received funding support from the NSRF via the Program Management Unit for Human Resources \& Institutional Development, Research and Innovation, grant B13F680083. CP is supported by Fundamental Fund 2569 of Khon Kaen University.

\appendix
\input{appendix}

\bibliographystyle{JHEP}
\bibliography{ref}

\end{document}

%% file: abstract.tex
\title{\boldmath Relocating the SIMP Miracle in the Axion Portal}

\abstract{%
In the strongly interacting massive particle (SIMP) scenario, dark matter is a pseudo-Nambu--Goldstone boson whose abundance is set by a three-to-two Wess--Zumino--Witten process.
We study this scenario in an axion portal and find that the kinetic contact the canonical mechanism assumes is excluded by bounds on sub-GeV axion-like particles (ALPs).
The dark sector then freezes out at its own temperature, so the relic abundance no longer fixes the self-interaction cross section but predicts the portal coupling instead.
Moreover, the portal operator is Hermitian and even in the ALP field, so no trilinear ALP--pion coupling arises.
The contact term then drives $\pi\pi\to aa$ with nothing to cancel against it, and we find the conversion four orders of magnitude faster than a trilinear estimate gives.
This disfavours the minimal realisation in which the dark condensate alone generates the ALP mass.
The realisation that survives instead makes the ALP slightly heavier than the dark pion. Matching the observed abundance then fixes the flavon vacuum expectation value at $V_\phi\simeq1.1\times10^{10}\GeV$ and leaves the dark scale open over more than an order of magnitude.
The decay $K^+\to\pi^+a$ requires the flavon to charge the leptons alone.
The model then predicts a dark matter self-interaction of $0.20\,{\rm cm^2/g}$ at a dark pion mass of $140\MeV$, in a window running from $82$ to $169\MeV$.
The predictive power of the SIMP framework is therefore not lost but relocated, from the self-interaction to the flavon scale.%
}

%% file: intro.tex

\section{Introduction}
\label{sec: intro}

The particle nature of dark matter (DM) remains the largest gap in our
description of the Universe~\cite{Bertone:2004pz,Planck:2018vyg}. For three
decades the weakly interacting massive particle (WIMP) has provided the
organising principle, since a particle with an electroweak-scale mass and an
electroweak-scale annihilation cross section freezes out of the primordial plasma
with approximately the observed abundance. The appeal of this ``WIMP miracle'' is
that a single measured number, $\Omega_{\rm DM}h^2=0.120\pm0.001$
\cite{Planck:2018vyg}, fixes a coupling that can be tested independently at
colliders and in direct-detection experiments. However, much of the WIMP
parameter space has now been excluded, and attention has therefore turned to the
sub-GeV thermal window, where the relic abundance is set not by $2\to2$
annihilation into Standard Model (SM) states but by number-changing processes
internal to the dark sector.

The strongly interacting massive particle (SIMP) scenario~\cite{Hochberg:2014dra}
is the sharpest realisation of this idea. If the dark sector is strongly coupled,
a $3\to2$ annihilation among the dark states can dominate over $2\to2$
annihilation and can by itself deplete the comoving number density to the
observed level. Because the $3\to2$ rate coefficient $\braket{\sigma v^2}_{3\to2}$ scales
as $\alpha_{\rm eff}^3/m_{\rm DM}^5$, matching the observed relic abundance drives the DM mass into the
$10$--$100\MeV$ range with a strongly coupled interaction. This is the regime in
which the DM self-interaction cross section per unit mass is of order
$\sigma_{\rm self}/m_{\rm DM}\sim0.1$--$1\,{\rm cm^2/g}$, the value long advocated
to alleviate the small-scale structure problems of collisionless cold dark
matter~\cite{Tulin:2017ara}. A concrete and calculable realisation is provided by
a QCD-like hidden gauge theory with dynamical chiral symmetry breaking, where the
dark pions are the DM and the $3\to2$ vertex descends from the
Wess--Zumino--Witten (WZW) term~\cite{Hochberg:2014kqa}. Matching the measured
relic abundance fixes the decay constant $f_\pi$, which leaves a tight correlation between
the relic abundance, the dark pion mass $m_\pi$ and the self-interaction cross
section. This correlation is what is usually called the \emph{SIMP miracle}. For
an $Sp(4)$ theory with $N_f=2$ it confines the viable window to roughly
$m_\pi\simeq75$--$400\MeV$, bounded from below by the bullet
cluster~\cite{Randall:2007ph} and from above by the perturbativity of the chiral
expansion, $m_\pi/f_\pi\lesssim2\pi$. The upper edge is sensitive to the WZW
group-theory factor $t^2$~\cite{Hochberg:2014kqa} and to the
$Sp(N_c)$ gauge rank, both of which enter $\braket{\sigma v^2}_{3\to2}$ as an
overall factor. For the minimal $N_f=2$ coset with $t^2=60$ and $N_c=2$, the upper
edge sits near $400\MeV$. Both edges are inherited from the assumption of kinetic
contact with the SM bath. Matching the relic abundance under that assumption fixes
$f_\pi$ once $m_\pi$ is chosen, so raising $m_\pi$ drives the ratio $m_\pi/f_\pi$ to
its perturbative limit.
It is this narrowness, and the attendant falsifiability, that makes the scenario
attractive. The mechanism has since been refined with unitarised chiral dynamics
and next-to-leading-order corrections~\cite{Cheng:2021kjg,Watanabe:2025zss}.

The SIMP miracle rests on an assumption that is often left implicit. Since the
$3\to2$ reaction converts rest mass into kinetic energy, a dark sector undergoing
it heats up relative to a freely expanding gas, and its temperature then falls
only logarithmically with the scale factor rather than as $a^{-2}$. This is the
cannibalism identified in Ref.~\cite{Carlson:1992fn}. Unless that entropy is
exported, the dark sector remains anomalously warm, structure formation is
modified, and the standard $3\to2$ freeze-out estimate does not apply, because it
assumes $T^\prime=T$. The original SIMP construction therefore requires a portal
that is strong enough to hold the dark pions in kinetic equilibrium with the SM
heat bath through freeze-out, and at the same time weak enough that $2\to2$
annihilation into SM states does not itself set the abundance. Threading this
needle is a genuine model-building constraint, and it is what forces the
introduction of a mediator, whether a kinetically mixed dark
photon~\cite{Hochberg:2014kqa}, a $Z^\prime$, or an axion-like particle
(ALP)~\cite{Kamada:2017tsq,Hochberg:2018rjs}. The tension is sharp enough that an
entire class of relics has been identified in the regime where the abundance is
fixed by the \emph{elastic} scattering rate on the SM rather than by $3\to2$ at
all, namely the elastically decoupling relics of
Refs.~\cite{Kuflik:2015bda,Kuflik:2017iqs}. The axion portal studied here does
not land in that regime, since the elastic energy exchange proceeds through the
off-shell $\pi f\to\pi f$ scattering, which is suppressed by $1/f_a^4$ and by the
small connector coupling and is therefore subdominant to $3\to2$ across the
viable portal range. In the opposite direction, it has been shown that if kinetic
contact with the SM is abandoned altogether, the SIMP survives as a consistent
self-interacting DM candidate whose viable parameter space migrates to larger
masses, where chiral perturbation theory is better
behaved~\cite{Heikinheimo:2018esa}. There the temperature ratio $\xi=T^\prime/T$
is treated as a free input and the freeze-out is rescaled accordingly. The step
we take here is to \emph{compute} $\xi$ from the portal coupling through the
energy-transfer equation, which turns the rescaling into a prediction rather than
a parametrisation.

Once the visible and dark sectors do not share a temperature, the dark sector
thermal evolution and relic density are governed by the temperature ratio
$\xi\equiv T^\prime/T$, whose value is fixed by the small portal coupling through
the energy that leaks into the hidden sector at early times. The resulting taxonomy
of relic-abundance regimes as the portal strength is varied at fixed hidden-sector
coupling, namely freeze-in, reannihilation, decoupled freeze-out and standard
freeze-out, has been mapped out in Ref.~\cite{Chu:2011be}. The thermodynamics of a
decoupled sector with a mass gap has been developed in
Refs.~\cite{Pappadopulo:2016pkp,Farina:2016llk}, where it has been emphasised
that such a sector generically enters a cannibal phase and that the resulting
warmness is constrained by the Lyman-$\alpha$ forest~\cite{Irsic:2017ixq} and by
$N_{\rm eff}$. Closely related dynamics appear in dark glueball scenarios, where
a pure-gauge hidden sector cannibalises through $3\to2$ glueball
interactions~\cite{Forestell:2016qhc,Soni:2016yes,Acharya:2017szw,Carenza:2022pjd,Carenza:2023eua}.
Most relevant for the present work, an ALP mediator between the SM and a dark
sector has been studied in Ref.~\cite{Bharucha:2022lty}, where an
intermediate-coupling regime called \emph{decoupled freeze-out} (DFO) has been
identified. In that regime the ALPs are produced abundantly enough by the SM bath
to thermalise the hidden sector among itself, but never equilibrate with the SM.
DFO interpolates between freeze-in and standard freeze-out, and it requires
solving a coupled system for $n_{\rm DM}$, $n_a$ and $T^\prime$.

The ALP is a particularly well-motivated connector. It is a
pseudo-Nambu--Goldstone boson, so its couplings to the SM are derivative or
mass-suppressed and technically natural over many decades. It couples to both
photons and fermions through
$\mathcal{L}\supset-\tfrac14 g_{a\gamma}aF\tilde F+i\sum_f(c_f/f_a)m_fa\bar
f\gamma_5f$~\cite{Wu:2024fsf}. Moreover, unlike a dark photon it couples naturally
to a confining dark sector through the same $e^{ia/f_a}\,{\rm Tr}\,J\Sigma$
structure that generates the dark pion mass, so a single operator controls $a\pi\pi$, $a^2\pi\pi$ and the ALP mass. It was in this spirit that the ``SIMP
through the axion portal'' was proposed in
Refs.~\cite{Kamada:2017tsq,Hochberg:2018rjs}. Since the same operator fixes the
ALP mass, the minimal choice in which no bare mass term is added ties $m_a$ to
the dark condensate, and we find that this tie is what ultimately decides the
fate of the scenario.

In this work, we map the axion-portal SIMP as a function of the ALP mass and the
portal coupling, and place it on the freeze-in/decoupled/freeze-out phase diagram
of Refs.~\cite{Chu:2011be,Bharucha:2022lty}. Two facts organise the map. First,
the WZW $3\to2$ process dominates the hidden-sector annihilation $\pi\pi\to aa$
whenever $f_a\gg f_\pi$, because that hierarchy makes the contact coupling
$g_{aa\pi\pi}=m_\pi^2/4f_a^2$ feeble. This holds for any ALP mass, so the thermal evolution is set by the portal
strength and not by the kinematics. For $m_a>m_\pi$ the channel $\pi\pi\to aa$ is
not closed but kinematically threshold-suppressed, and at $m_a=1.1\,m_\pi$ the
suppression at freeze-out is only $e^{-0.1x}\sim e^{-2}$. The earlier claim that
$m_a>m_\pi$ is required, and that $\pi\pi\to aa$ overtakes $3\to2$ by many orders
as soon as it opens, is therefore too strong and we drop it. Second, kinetic
contact with the SM does not require the on-shell ALP. It is carried by elastic
$\pi a\to\pi a$ scattering, enhanced for $m_a<m_\pi$ by
$n_a/n_\pi\sim e^{(m_\pi-m_a)/T^\prime}$, and by off-shell $\pi f\to\pi f$
scattering on SM fermions, which carries no ALP-mass suppression and so persists
for $m_a>m_\pi$. We compute both explicitly rather than argue them away. The model
then has two branches. In one, contact survives, $3\to2$ sets the abundance at
$T^\prime=T$, and the ordinary SIMP miracle holds. In the other the portal is
feeble, the sector decouples, and it cannibalises at its own temperature
$T^\prime=\xi T$. What separates them is the portal strength and not the ALP mass,
and the divide is a band rather than a line, since the sectors reach a common
temperature by energy transfer before elastic scattering can maintain it through
freeze-out. We find that the contact branch requires an ALP--SM coupling
$g_{aff}=c_f/f_a$ in the range already probed by beam-dump and rare-decay searches
for a sub-GeV electrophilic ALP, so once the portal is weak enough to be viable
the sector decouples for any ALP mass. Two related constructions sharpen the
point. In Ref.~\cite{Fiorentino:2026} the same electron-coupled ALP is studied at
$m_a\gtrsim m_\pi$ with contact held through a non-zero $\theta$ angle, and that
$\theta$ angle drives $3\to2$ through cubic pion couplings independent of the WZW
term~\cite{GarciaCely:2024}. That is a different mechanism rather than a
counterexample.

We then work out the consequences in the decoupled branch, where the observed
relic sits. Kinetic equilibrium within the dark sector is maintained not by the
ALP but by dark pion self-scattering, which stays faster than Hubble at
freeze-out and is guaranteed by the very self-interaction strength that the SIMP
mechanism requires. The dark sector is then a cannibal gas whose comoving entropy
grows while the portal is injecting energy and is conserved afterwards, and the
relic yield follows the scaling
$\Omega_\pi h^2 \propto (s^\prime/s)\,T^\prime_f$, where $s^\prime/s$ is the
injected entropy per unit visible entropy, $T^\prime_f$ is the dark freeze-out
temperature and the $3\to2$ coupling enters only logarithmically.
We verify this numerically and find that $\Omega_\pi h^2$ is nearly independent of
$m_\pi/f_\pi$. The relic abundance therefore no longer
determines $f_\pi$, and the chain $\Omega h^2\to f_\pi\to\sigma_{\rm self}/m_\pi$
no longer holds. The SIMP miracle in its usual form does not survive. The
self-interaction cross section becomes a free parameter, bounded only by the bullet
cluster and by perturbativity, and the narrow SIMP mass window opens up to the GeV scale.
This recovers, from an explicit portal, the larger-mass self-interacting window
anticipated in Ref.~\cite{Heikinheimo:2018esa}, where it followed from treating
$\xi$ as a free input. Here it follows from an injection that the portal fixes.

This loss of predictivity is only apparent, and recovering it is the second point
of this work. The injected entropy is not a free input. It is generated by the
energy that the ALP portal leaks from the SM into the hidden sector, and is
therefore a calculable function of $g_{aff}$. Which power of $g_{aff}$ appears
depends on whether the inverse decay $f\bar f\to a$ is open. When it is, the
injected energy density scales as $g_{aff}^2$, and since that energy arrives after
the dark bath has become non-relativistic the entropy it deposits is linear in it,
so $\Omega_\pi h^2\propto g_{aff}^2$ up to a logarithm. When the inverse decay is
closed the injection proceeds through $f\bar{f}\to aa$ transfer and the energy density
scales as $g_{aff}^4$ instead. Either way the relic abundance constraint becomes a
condition on the portal coupling. Closing that loop
converts the statement ``$\Omega_\pi h^2=0.12$ fixes the injected entropy'' into
``$\Omega_\pi h^2=0.12$ fixes $g_{aff}$''. The miracle is not lost but
\emph{relocated}. Instead of predicting a DM self-interaction cross section, which
is measurable only through astrophysical systematics, the observed relic abundance
now predicts an ALP--SM coupling.

The minimal realisation then decides the character of the scenario, and it does
so through a single structural feature of the portal. The operator carries its
Hermitian conjugate, and the invariant it contains is real, so the portal is even
in the ALP field. No trilinear $a\pi\pi$ coupling is generated, the dark sector
couples to the ALP only in pairs, and the conversion $\pi\pi\to aa$ proceeds
through a contact term with no near-cancellation to slow it down. We find that
this excludes the most economical version of the model. If the dark condensate
alone generates the ALP mass, then $m_a=\sqrt{N_f}\,m_\pi f_\pi/2f_a$ places the
ALP two orders below the dark pion, and the conversion carries no threshold. The
requirement that the $3\to2$ process dominate then pushes the dark scale above the
ceiling that supernova ALPs impose, at every dark pion mass.

The realisation that survives keeps the flavon but abandons the tie. We identify 
the ALP with the axiflavon, whose vacuum expectation value $V_\phi$ sets the
couplings to SM fermions and photons through Froggatt--Nielsen charges, while
$f_a$ remains the dark scale that enters the portal to the pions. Once the ALP is
slightly heavier than the dark pion the conversion is Boltzmann suppressed at
freeze-out, and we find that $m_a\simeq1.25\,m_\pi$ removes the floor entirely.
Moreover $aa\to\pi\pi$ is then exothermic, so the ALP annihilates into dark pions
rather than surviving as a hot relic, and the spectral distortion and
photodissociation bounds that would otherwise apply have nothing to act on. The
energy that the portal leaks no longer runs through a trilinear vertex, so it does
not carry the dark scale, and matching the observed abundance fixes the flavon
scale outright,
\begin{equation}
  V_\phi \simeq 1.1\times10^{10}\GeV ,
  \label{eq: intro-vphi}
\end{equation}
while leaving $f_a$ free. The decay $K^+\to\pi^+a$ then decides which fermions
the flavon may charge. A flavon carrying the quark hierarchy predicts a branching
ratio one to two orders above the NA62 limit on $K^+\to\pi^+X$, so the flavon
must charge the leptons alone. The ALPs a supernova produces do leave
the star at this scale, but Primakoff conversion carries two powers of the photon
coupling, so the yield falls as $V_\phi^{-2}$ and lands five orders below what the
Galactic $511\keV$ line allows~\cite{Jean:2005af}.

What tests the model instead is the dark sector itself. Since the relic no longer
fixes the self-interaction cross section, that quantity is set by the dark pion
mass alone, and at $m_\pi=140\MeV$ we find
$\sigma_{\rm self}/m_\pi\simeq0.20\,{\rm cm^2/g}$, inside the
range invoked to address the small-scale structure of haloes. The bullet cluster bounds it from
above and the dimuon threshold bounds the dark pion mass from the other side. The
surviving window is therefore narrow, and it could be probed by halo shapes and
cluster mergers rather than by MeV astronomy.

The paper is organised as follows. Section~\ref{sec: model} defines the model,
where we introduce the $Sp(N_c)$ hidden sector, its chiral Lagrangian and WZW
term, the ALP portal to the dark pions, and the ALP couplings to SM fermions and
photons, together with the parameter set
$(m_\pi,f_\pi,m_a,f_a,V_\phi,E)$. Section~\ref{sec: regimes} classifies the
production regimes as a function of the connector coupling, from freeze-in through
decoupled freeze-out to standard freeze-out, and sets up the coupled Boltzmann
system for $(n_\pi,n_a,T^\prime)$ including the energy-transfer collision integral
that determines $\xi$. Section~\ref{sec: consistency} maps the two branches in the
plane of $m_a/m_\pi$ and the portal coupling, locating the $3\to2$ domination
boundary and the kinetic-decoupling boundary. Section~\ref{sec: cannibal DFO}
solves the cannibal freeze-out in the decoupled sector and derives the
$\Omega h^2\propto(s^\prime/s)\,T^\prime_f$ scaling. Section~\ref{sec: relocated}
closes the loop by computing the injected entropy from the energy-transfer
equation.
Section~\ref{sec: tied} imposes the minimal realisation, disfavours the
condensate-tied mass, works out the branch that survives, and follows the ALPs
that a core-collapse supernova produces. Section~\ref{sec: pheno} collects the remaining astrophysical
and laboratory constraints. We finally conclude in Sec.~\ref{sec: conclusions}, and technical
material is collected in the appendices, covering the cross sections and squared
amplitudes, the thermal averages, the collision integrals and the numerical
implementation.

%% file: sec_model.tex

\section{The model}
\label{sec: model}

\begin{figure}[t]
  \centering
  \includegraphics[width=1\linewidth]{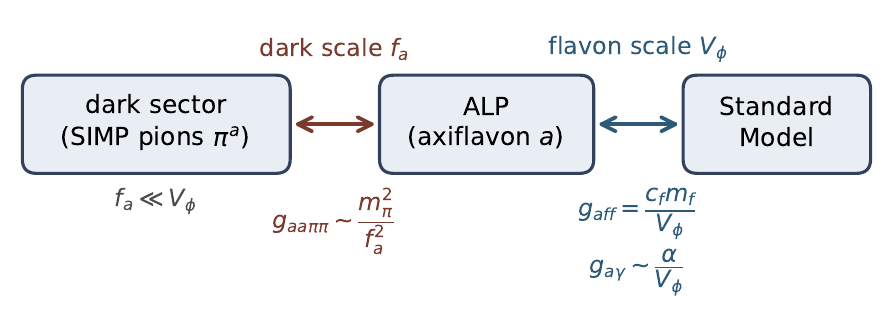}
  \caption{The three sectors of the model and their two scales. The axion-like
  particle couples to the dark pions on the dark scale $f_a$ and to the Standard
  Model (SM) fermions and photons on the flavon scale $V_\phi$, with $f_a\ll V_\phi$.
  There is no direct coupling between the dark pions and the SM.}
  \label{fig: three sectors}
\end{figure}

The model consists of three sectors. The first is a confining dark sector whose lightest
states, the dark pions, are the dark matter. The second is an axion-like particle
(ALP), which acts as a mediator. The third is the SM.
The symmetry structure of the theory is organised as follows:
\begin{enumerate}[label=(\roman*)]
  \item \textbf{Dark gauge symmetry:} A confining $Sp(N_c)$ gauge group with confinement scale $\Lambda_D$.
  \item \textbf{Dark flavour symmetry:} A global $SU(2N_f)$ flavour symmetry under which the dark quarks transform, broken dynamically to $Sp(2N_f)$ by the dark quark condensate.
  \item \textbf{Peccei--Quinn (PQ) / Froggatt--Nielsen symmetry:} A global $U(1)_{\rm PQ}$ symmetry spontaneously broken at the flavon scale $V_\phi$ in the visible sector, and nonlinearly realised at the scale $f_a$ in the dark sector.
  \item \textbf{SM gauge symmetry:} The standard $SU(3)_{\rm c} \times SU(2)_{\rm L} \times U(1)_{\rm Y}$ gauge interactions.
\end{enumerate}
The ALP is the sole connector between the hidden and visible sectors, coupling to the dark pions through the mass-generating operator and to the SM through the usual ALP couplings to fermions and gauge bosons, while the dark pions and the SM do not couple directly. Figure~\ref{fig: three sectors} summarises this structure.

\subsection{The hidden sector and the dark pions}

The dark sector is an $Sp(N_c)$ gauge theory with $N_f$ Dirac fermions, the dark
quarks (DQs) $Q_{D,i}$, in the fundamental representation,
\begin{equation}
  \mathcal{L}_{\rm dark}
  = -\tfrac14 G^A_{\mu\nu}G^{A\,\mu\nu}
    + \sum_{i=1}^{N_f}\bar Q_{D,i}\,\big(i\slashed{D} - m_Q\big)\,Q_{D,i} ,
  \label{eq: gauge}
\end{equation}
where $G^A_{\mu\nu}$ is the gauge field strength, $D_\mu$ the covariant derivative
and $m_Q$ a common DQ mass. Because the fundamental of $Sp(N_c)$ is
pseudoreal, the $2N_f$ Weyl components carry an enlarged global $SU(2N_f)$ flavour
symmetry rather than the $SU(N_f)_L\times SU(N_f)_R$ of a QCD-like
theory~\cite{Hochberg:2014kqa,Hochberg:2018rjs}. Confinement forms a quark
condensate $\braket{\bar Q_D Q_D}\simeq\mu^3$, where $\mu$ is the dark
chiral-symmetry-breaking scale, and this condensate breaks $SU(2N_f)$ to
$Sp(2N_f)$, leaving
\begin{equation}
  N_\pi = (N_f-1)(2N_f+1)
  \label{eq: Npi}
\end{equation}
pseudo-Nambu--Goldstone bosons, the dark pions $\pi^a$. They are collected into
the matrix $\Sigma=\exp(2i\pi/f_\pi)\,J$, with $\pi=\pi^a X^a$, where $f_\pi$ is
the dark pion decay constant, $X^a$ are the broken generators and $J$ is the antisymmetric
$Sp(2N_f)$ invariant. Their dynamics follow from the leading chiral Lagrangian~\cite{Witten:1983tx,Hochberg:2014kqa,Kosower:1984aw}
\begin{equation}
  \mathcal{L}_{\rm ChPT}
  = \frac{f_\pi^2}{4}\,{\rm Tr}\big[\partial_\mu\Sigma^\dagger\partial^\mu\Sigma\big]
    - \frac{\mu^3 m_Q}{2}\,{\rm Tr}[J\Sigma] + {\rm h.c.} ,
  \label{eq: chiral}
\end{equation}
where the second term is the DQ mass insertion, which gives the dark pions a
common mass
\begin{equation}
  m_\pi^2 = \frac{8 m_Q\mu^3}{f_\pi^2} ,
  \label{eq: gmor}
\end{equation}
the dark analogue of the Gell-Mann--Oakes--Renner relation~\cite{Gell-Mann:1968hlm}, where the Hermitian
conjugate doubles the surviving coefficient and ${\rm Tr}[\pi^2]=2\pi^a\pi^a$ in
the convention of Appendix~\ref{app: portal}. The sign of the mass insertion is
fixed by ${\rm Tr}[J\Sigma]=-{\rm Tr}[e^{2i\pi/f_\pi}]$, which is what makes the
pion potential stable, and it is the $a\to0$ limit of
Eq.~\eqref{eq: portal-op}. The chiral expansion is
valid for $m_\pi/f_\pi\lesssim2\pi$, which sets the upper edge of the viable mass
window.

The leading number-changing interaction among the pions is the
Wess--Zumino--Witten (WZW) term~\cite{Wess:1971yu,Witten:1983tx,Witten:1983tw}
\begin{equation}
  \mathcal{L}_{\rm WZW}
  = \frac{2 N_c}{15\pi^2 f_\pi^5}\,\epsilon^{\mu\nu\rho\sigma}
    \sum_{i<j<k<l<m} T_{ijklm}\,
    \pi_i\,\partial_\mu\pi_j\,\partial_\nu\pi_k\,\partial_\rho\pi_l\,
    \partial_\sigma\pi_m ,
  \label{eq:WZW term}
\end{equation}
where $T_{ijklm}$ is the totally antisymmetric group-theory tensor of the coset. It
drives the $3\to2$ annihilation that sets the relic abundance, with the thermally
averaged rate~\cite{Hochberg:2014kqa,Hochberg:2014dra}
\begin{equation}
  \braket{\sigma v^2}_{3\to2}
  = \frac{5\sqrt5}{2\pi^5 x_f^{\prime\,2}}\,
    \frac{N_c^2\, m_\pi^5}{f_\pi^{10}}\,\frac{t^2}{N_\pi^3} ,
  \label{eq: sv2}
\end{equation}
where $x_f^\prime=m_\pi/T_f^\prime\simeq20$ is the freeze-out variable evaluated at
the dark-sector temperature $T_f^\prime$ and
$t^2=\tfrac23N_f(N_f^2-1)(4N_f^2-1)$ is the group-theory factor. The topological
Wess--Zumino--Witten term exists only for $N_f\ge2$~\cite{Wess:1971yu,Witten:1983tx,Witten:1983tw}, and $N_f=2$ is therefore the
minimal realisation of the $3\to2$ mechanism, on which we focus. The gauge rank
enters only through the prefactor $N_c^2$. We keep $N_c$, $N_f$ and $N_\pi$ general
throughout the model, and specialise to $N_f=2$, for which the coset is
$SU(4)/Sp(4)$ with $N_\pi=5$ and $t^2=60$, and to $N_c=2$ for benchmark numerical results from Sec.~\ref{sec: regimes} onwards.

For the $N_f=2$ coset the cubic pion invariant vanishes, ${\rm Tr}[\pi^3]=0$, so no
cubic pion vertex is generated at this order. A non-zero $\theta$ angle would
induce such a vertex and open a second number-changing
channel~\cite{GarciaCely:2024}. We work at $\theta=0$, where the WZW term is the
only number-changing interaction in the dark sector.

The stability of the dark pions, which is what makes them viable dark matter,
follows from an accidental symmetry. The pions transform in a real representation of
the unbroken $Sp(2N_f)$ flavour symmetry, and every interaction is a flavour
singlet. Since the SM and the ALP are flavour singlets, they couple to
the pions only through the invariant bilinear $\pi^a\pi^a$, so the dark pions are
produced from and annihilate into the visible sector only in pairs. This acts as an
accidental $\mathbb{Z}_2$ parity under which the pions are odd, so a single dark pion
has no decay channel to the SM, and being the lightest state of a
conserved flavour multiplet it is stable~\cite{Hochberg:2014kqa}. The internal WZW
$3\to2$ reaction rearranges pions within the dark sector but does not spoil this,
since it too is a flavour singlet.

\subsection{The axion-like particle and its portal to the dark pions}

The ALP couples to the dark pions through the same spurion structure that
generates the pion mass, so that the portal Lagrangian reads
\begin{equation}
  \mathcal{L}_{a\pi}
  = -\tfrac12 m_{\rm PQ}^2 a^2
    - \tfrac12 m_Q\mu^3\, e^{ia/f_a}\,{\rm Tr}[J\Sigma] + {\rm h.c.} ,
  \label{eq: portal-op}
\end{equation}
where $f_a$ is the ALP decay constant and $m_{\rm PQ}$ is a bare mass
from explicit PQ breaking above the condensate scale. Adding the Hermitian
conjugate makes the operator even in $a$ and even in $\pi$ separately, so every
vertex with an odd number of ALP legs cancels. This is not a choice of vacuum.
A constant misalignment of the ALP field can always be removed by the shift
$a\to a-f_a\bar\theta$~\cite{Peccei:1977hh,Peccei:1977ur}, so an odd vertex survives only if the ALP potential is
minimised away from $a=0$, and here it is not. Both sources of Peccei--Quinn
breaking, the portal term of Eq.~\eqref{eq: portal-op} and the bare mass
$m_{\rm PQ}^2$, are taken to be minimised at $a=0$. We assume this alignment
rather than derive it. A generic explicit-breaking potential is a cosine carrying
its own phase, and the bare term is its expansion about its own minimum, so the
two are aligned only if the ultraviolet breaking conserves CP in the same basis as
the condensate. Granting that, no trilinear $a\pi\pi$ coupling exists in either of
the two cases below, and generating one would require a second breaking term
misaligned with the portal. We derive this in Sec.~\ref{sec: tied} and find that it excludes the
minimal realisation.
Expanding in powers of $\pi/f_\pi$ and $a/f_a$ then leaves
\begin{align}
  \mathcal{L}_{a\pi} \supset\;
  & -\tfrac12\Big(m_{\rm PQ}^2 + \frac{N_f\,m_\pi^2 f_\pi^2}{4 f_a^2}\Big)a^2
    - \tfrac12 m_\pi^2\,\pi^a\pi^a
    + \frac{1}{4}\frac{m_\pi^2}{f_a^2}\,a^2\,\pi^a\pi^a ,
  \label{eq:ALP-DM L}
\end{align}
with the full expansion given in Appendix~\ref{app: portal}. Three features matter
for what follows. First, there is no $a\pi\pi$ vertex, and none of the odd
couplings $a^3$ and $a^3\pi\pi$ that a single exponential would generate. The
leading portal interaction is instead the quartic contact term
\begin{equation}
  g_{a a\pi\pi} = \frac{m_\pi^2}{4 f_a^2} ,
\end{equation}
which alone controls the hidden annihilation $\pi\pi\to aa$ and the elastic
scattering $\pi a\to\pi a$. Second, the same single operator fixes the $a^2$ and
$a^2\pi\pi$ couplings, so the dark-sector phenomenology is controlled by
$f_a$ alone once $m_\pi$ and $f_\pi$ are given. Third, the ALP mass receives two contributions, so from
Eq.~\eqref{eq:ALP-DM L} it reads
\begin{equation}
  m_a^2 = m_{\rm PQ}^2 + \frac{N_f\,m_\pi^2 f_\pi^2}{4 f_a^2} ,
  \label{eq: tied-mass}
\end{equation}
the first from explicit PQ breaking above the condensate scale and the
second from the dark condensate itself. How these compete is a genuine
model-building choice rather than a technicality, and it distinguishes two cases.
In the \emph{generic} case the bare term dominates, $m_a\simeq m_{\rm PQ}$, so $m_a$
is a free parameter and the condensate piece is a negligible correction for
$f_a\gg f_\pi$. Sections~\ref{sec: regimes}--\ref{sec: relocated} work in this case,
because it keeps $m_a$ and $f_a$ independent and allows the phase diagram to be
scanned in both. In the \emph{minimal} case the shift symmetry is broken only by
the dark condensate, $m_{\rm PQ}=0$, and the ALP mass is generated entirely by the
portal,
\begin{equation}
  m_a = \frac{\sqrt{N_f}\,m_\pi f_\pi}{2 f_a} ,
  \label{eq: tied-min}
\end{equation}
so that one of $(f_\pi,m_a)$ becomes redundant. This is the tied-mass constraint.
It is the most economical version of the model, since it introduces no scale beyond
those already present, and it is therefore the limit to test first. It places the
ALP two orders below the dark pion, and we find in Sec.~\ref{sec: tied} that this
is what excludes it, because the conversion $\pi\pi\to aa$ then carries no
threshold. The generic case survives, and the ALP mass there is a free parameter
which the same section fixes just above $m_\pi$. When quoting numbers we take $N_f=2$, so that
$m_a=\sqrt2\,m_\pi f_\pi/2f_a$, and treat the remaining $\mathcal{O}(1)$ coset
ambiguity as part of the residual mass-window uncertainty.

\subsection{The axiflavon and its couplings to the SM}

We identify the ALP with the axiflavon, the phase of a complex scalar $\Phi$ that
is an SM singlet and carries charge under a global $U(1)_{\rm PQ}$
symmetry~\cite{Ema:2016ops,Calibbi:2016hwq}. Writing the $\Phi$ in terms of its
modulus and phase,
\begin{equation}
  \Phi = \left(\frac{V_\phi + \phi}{\sqrt2}\right)\,e^{i a/V_\phi} ,
  \label{eq: flavon}
\end{equation}
where $V_\phi=\sqrt2\,\braket{\Phi}$ is the vacuum expectation value. The field $\Phi$ contains two excitations, the CP-even flavon $\phi$ and the CP-odd axiflavon $a$, following Ref.~\cite{Calibbi:2016hwq}. The flavon acquires a heavy mass $m_\phi\sim V_\phi\sim10^{10}\GeV$ upon symmetry breaking and is safely integrated out well above the temperatures relevant for freeze-out, leaving the axiflavon as the active low-energy degree of freedom. We call it the ALP in what follows, since the constraints of Sec.~\ref{sec: pheno} are the generic ALP ones. The decay constant of the phase is the vacuum expectation value itself, so under $U(1)_{\rm PQ}$ the axiflavon shifts as $a\to a + V_\phi\,\alpha$, where $\alpha$ is the constant transformation parameter of the global symmetry. The same field generates the Yukawa hierarchy through the Froggatt--Nielsen mechanism~\cite{Froggatt:1978nt,Calibbi:2016hwq,Greljo:2024evt}, the SM Yukawa operators arising as
\begin{equation}
  \mathcal{L}_Y = c_{ij}\Big(\frac{\Phi}{\Lambda}\Big)^{n_{ij}}\,
    \bar\psi_i\, H\, \psi_j + {\rm h.c.} ,
  \label{eq: FN-yukawa}
\end{equation}
where $\Lambda$ is the flavour scale, the $c_{ij}$ are $\mathcal{O}(1)$ and the
exponents $n_{ij}=X_{\psi_i}+X_{\psi_j}$ are the sums of the Froggatt--Nielsen
charges of the two fields, written as left-handed Weyl fermions, so that after
$\Phi$ acquires its vacuum expectation value the physical Yukawa is
$y_{ij}=c_{ij}\,\epsilon^{n_{ij}}$ with the Froggatt--Nielsen suppression
$\epsilon=V_\phi/\sqrt2\Lambda\simeq0.22$ fixed to the Cabibbo
angle~\cite{Froggatt:1978nt,Leurer:1992wg,Calibbi:2016hwq}.
The flavour scale is therefore $\Lambda\simeq V_\phi/\sqrt2\epsilon\simeq3\,V_\phi$,
a few times the flavon vacuum expectation value. We do not attempt to reproduce the
full quark and lepton mass and mixing structure, which is the subject of the
axiflavon literature~\cite{Calibbi:2016hwq}, and take the charges as $\mathcal{O}(1)$
inputs. We find in Sec.~\ref{sec: tied} that the quark charges are in any case
forced to vanish once the supernova constraint is imposed, so the flavon of this
paper carries the charged-lepton hierarchy alone.

Expanding the phase in Eq.~\eqref{eq: flavon} and rotating to the fermion mass basis
gives the diagonal ALP couplings
\begin{equation}
  \mathcal{L}_{af}
  = \frac{\partial_\mu a}{V_\phi}\sum_f c_f\,\bar f\gamma^\mu\gamma_5 f
  \;\supset\; i\sum_f \frac{c_f\, m_f}{V_\phi}\,a\,\bar f\gamma_5 f ,
  \label{eq: ALP-SM L}
\end{equation}
where the visible decay constant is the flavon vacuum expectation value $V_\phi$ and
\begin{equation}
  c_f = X_{f_L} + X_{f_R}
  \label{eq: axiflavon}
\end{equation}
is the sum of the Froggatt--Nielsen charges of the left-handed doublet and the
right-handed singlet, an $\mathcal{O}(1)$ integer whose magnitude is the exponent
$n_f$ that reproduces the Yukawa $y_f\sim\epsilon^{|c_f|}$. A fermion the flavon
does not charge keeps $c_f=0$ and decouples from the ALP entirely, its Yukawa coming
from elsewhere. Under the PQ symmetry the
axion shifts as $a\to a+V_\phi\,\alpha$ while the fermions rotate as
$f_L\to e^{iX_{f_L}\alpha}f_L$ and $f_R\to e^{iX_{f_R}\alpha}f_R$, where $\alpha$ is the global transformation parameter, so that the
derivative coupling of Eq.~\eqref{eq: ALP-SM L} is invariant, the shift of the
axion being compensated by the chiral rotation of the fermions. Within a charged
sector the lighter a fermion the larger its charge, so the electron carries the
largest charge of all, a pattern that controls both the energy injection and the
photon coupling. Which fermions the flavon charges is not a free choice once the
astrophysical constraints are imposed, and we find in Sec.~\ref{sec: tied} that it
must charge the leptons alone.

Since the ALP couples to matter only through the charges of
Eq.~\eqref{eq: axiflavon}, its coupling to photons is not an independent operator
but is generated by the fermion triangle, so that it is the electromagnetic anomaly
of the $U(1)_{\rm PQ}$,
\begin{equation}
  g_{a\gamma}
  = \frac{\alpha_{\rm em}}{2\pi V_\phi}\sum_f c_f\, N_c^f\, Q_f^2
  \equiv \frac{\alpha_{\rm em}}{2\pi V_\phi}\,E ,
  \label{eq: gag-total}
\end{equation}
where the sum runs over every PQ-charged fermion and $E$ is the anomaly
coefficient, an $\mathcal{O}(1)$--$\mathcal{O}(10)$ integer~\cite{Calibbi:2016hwq,Wu:2024fsf}.
The charged SM fermions give a definite contribution to $E$, computed in
Sec.~\ref{sec: tied}, and any heavier PQ-charged states above $\Lambda$
add to it, so we take $E$ as a model input bounded from below by its SM
part. The companion colour anomaly $N=\tfrac12\sum_q c_q$ would give the ALP a
coupling to gluons, and hence to nucleons, and Sec.~\ref{sec: tied} shows that it
has to vanish. There is no separate tree-level $a\gamma\gamma$ term.

The visible couplings of Eqs.~\eqref{eq: ALP-SM L} and \eqref{eq: gag-total} run on
the flavon scale $V_\phi$, while the portal to the dark pions of
Eq.~\eqref{eq: portal-op} runs on the dark scale $f_a$. The two are connected by the
DQs, which also carry PQ charge, so that the dark condensate
depends on the axion phase and generates both the tied mass of
Eq.~\eqref{eq: tied-min} and the contact coupling $g_{aa\pi\pi}$. The axion does not mix linearly
with a single dark pion, since the diagonal invariant ${\rm Tr}[\pi]$ vanishes in
the $Sp(N_c)$ coset, so the mass arises directly from the $a^2$ term of
Eq.~\eqref{eq:ALP-DM L} rather than from an axion--$\pi_0$ mixing as for the QCD
axion. The effective dark decay constant is then $f_a=V_\phi/X_Q$, with $X_Q$ the
PQ charge of the DQ mass operator, so the hierarchy $f_a\ll V_\phi$ needed for a
decoupled freeze-out would require $X_Q\sim10^9$. We do not attempt to generate a
charge of that size. At the level of the effective Lagrangian of
Eq.~\eqref{eq: portal-op}, however, $f_a$ is an independent parameter, and we take
$f_a$ and $V_\phi$ as independent inputs, leaving a complete ultraviolet embedding
to future work. In the generic case of Sections~\ref{sec: regimes}--\ref{sec: relocated}
the two scales coincide, $V_\phi=f_a$.

\subsection{Parameters}

The model is fixed by the parameters
\begin{equation}
  (m_\pi,\; f_\pi,\; m_a,\; f_a,\; V_\phi,\; E) ,
\end{equation}
with the dimensionless couplings $c_f m_f/V_\phi$ and the photon coupling
$g_{a\gamma}=(\alpha_{\rm em}/2\pi V_\phi)E$ both set by the axiflavon charges of
Eqs.~\eqref{eq: axiflavon} and \eqref{eq: gag-total}. In the generic case
$V_\phi=f_a$ and the two scales collapse to one. The dark pion mass $m_\pi$ and
decay constant $f_\pi$ set the strength of the $3\to2$ process through
Eq.~\eqref{eq: sv2} and the self-interaction cross section. The dark scale $f_a$
sets the portal to the dark pions through $g_{aa\pi\pi}$, and the flavon scale
$V_\phi$ together with the anomaly coefficient $E$ sets the portal to the SM. The ALP mass $m_a$ enters through the kinematics of $\pi\pi\to aa$ and
$\pi a\to\pi a$ and through the
ALP lifetime. In the minimal case the tied-mass constraint of
Eq.~\eqref{eq: tied-min} removes one of these. The rest of the paper studies the dark sector thermal evolution as a function
of the hidden-sector strength $m_\pi/f_\pi$ and the connector coupling $g_{aff}$,
at fixed $m_\pi$ and $m_a$.

%% file: sec_regimes.tex

\section{Production regimes and the Boltzmann system}
\label{sec: regimes}

The relic density of the axion-portal SIMP is controlled by three couplings. The first
is the strength of the Wess--Zumino--Witten $3\to2$ vertex, which is set by the
pion decay constant and enters the thermally averaged rate as
$\braket{\sigma v^2}_{3\to2}\propto m_\pi^5/f_\pi^{10}$. The second is the
quartic contact coupling $g_{aa\pi\pi}=m_\pi^2/4f_a^2$, which controls the
hidden-sector annihilation $\pi\pi\to aa$ and the elastic scattering
$\pi a\to\pi a$, and which descends from the mass term. It is the leading portal
interaction, since Sec.~\ref{sec: model} leaves no trilinear $a\pi\pi$ vertex. The third is the ALP--fermion coupling
$g_{aff}=c_f/V_\phi$, the derivative coupling of Eq.~\eqref{eq: ALP-SM L} in
inverse GeV, which controls the exchange with the SM and is set
by the flavon scale $V_\phi$ of Sec.~\ref{sec: model}. In the generic case treated in
Sections~\ref{sec: regimes}--\ref{sec: relocated} the dark and visible scales coincide,
$V_\phi=f_a$, so that $g_{aff}=c_f/f_a$, and the portal to the pions shares this decay
constant. For a fixed $c_f$ the two move together, so a single knob $f_a$ sets the
whole portal.

This is the one structural difference from the ALP-mediator study of
Ref.~\cite{Bharucha:2022lty}. There the dark matter number is changed by a
portal-mediated $2\to2$ annihilation into ALPs, so the hidden coupling and the
relic-setting coupling are the same object. Here the relic is set by the $3\to2$
process, which does not involve the ALP at all, so the hidden number-changer is
decoupled from the portal. We find below that this single fact rotates the
relic-density contour in the phase diagram, and it is the origin of the relocated
miracle.

\subsection{The coupled Boltzmann system}

We work throughout in the regime where the portal couplings are small, so that the
dark sector never reaches thermal equilibrium with the SM. The two
sectors are then generically populated separately in the early Universe and carry
their own temperatures, $T$ for the SM and $T^\prime$ for the dark
sector, and the portal only leaks a small amount of energy from one to the other.
The dark sector is described by three quantities, the pion number density $n_\pi$,
the ALP number density $n_a$ and the dark temperature $T^\prime$, which obey
\begin{align}
  \frac{d n_\pi}{dt} + 3H n_\pi
  &= -\braket{\sigma v^2}_{3\to2}\big(n_\pi^3 - n_\pi^2\, n_\pi^{\rm eq\prime}\big)
     -\braket{\sigma v}_{\pi\pi\to aa}\big(n_\pi^2 - n_a^2\,\mathcal{R}\big)
     + S_\pi ,
  \label{eq: boltz-npi}\\
  \frac{d n_a}{dt} + 3H n_a
  &= +\braket{\sigma v}_{\pi\pi\to aa}\big(n_\pi^2 - n_a^2\,\mathcal{R}\big)
     -\braket{\Gamma_a}\big(n_a - n_a^{\rm eq}\big)
     + S_a ,
  \label{eq: boltz-na}\\
  \frac{d\rho^\prime}{dt} + 3H(\rho^\prime + P^\prime)
  &= C_E ,
  \label{eq: boltz-T}
\end{align}
where $\mathcal{R}=(n_\pi^{\rm eq\prime}/n_a^{\rm eq\prime})^2$ enforces detailed
balance and the equilibrium densities marked with a prime are evaluated at
$T^\prime$. The source terms $S_\pi$ and $S_a$ are the production of pions and ALPs
from the SM bath through the portal, fixed by detailed balance at the
visible temperature $T$,
\begin{equation}
  S_\pi = \sum_f\braket{\sigma v}_{\pi\pi\to f\bar f}\big[(n_\pi^{\rm eq})^2 - n_\pi^2\big](T) ,
  \qquad
  S_a   = \sum_f\braket{\sigma v}_{aa\to f\bar f}\big[(n_a^{\rm eq})^2 - n_a^2\big](T) ,
  \label{eq: sources}
\end{equation}
where the unprimed equilibrium densities $n_\pi^{\rm eq}$ and $n_a^{\rm eq}$ are
evaluated at the SM temperature $T$, so that each source drives its
species towards the visible-bath equilibrium and vanishes there, and the sum runs
over the kinematically accessible SM fermions. The first
equation carries the number-changing $3\to2$ reaction and the conversion
$\pi\pi\leftrightarrow aa$, and the second carries the same conversion with the
opposite sign together with the ALP decay and inverse decay. The third equation is
the energy balance of the dark sector, whose source $C_E$ is the energy
transferred per unit time and volume from the SM, from inverse decays and from $f\bar f\to aa$. It is this term that fixes the
temperature ratio $\xi\equiv T^\prime/T$. The collision integrals and the
energy-transfer term are given in Appendix~\ref{app: collision}.

\subsection{The phase diagram}

\begin{figure}[p]
  \centering
  \includegraphics[width=0.86\columnwidth]{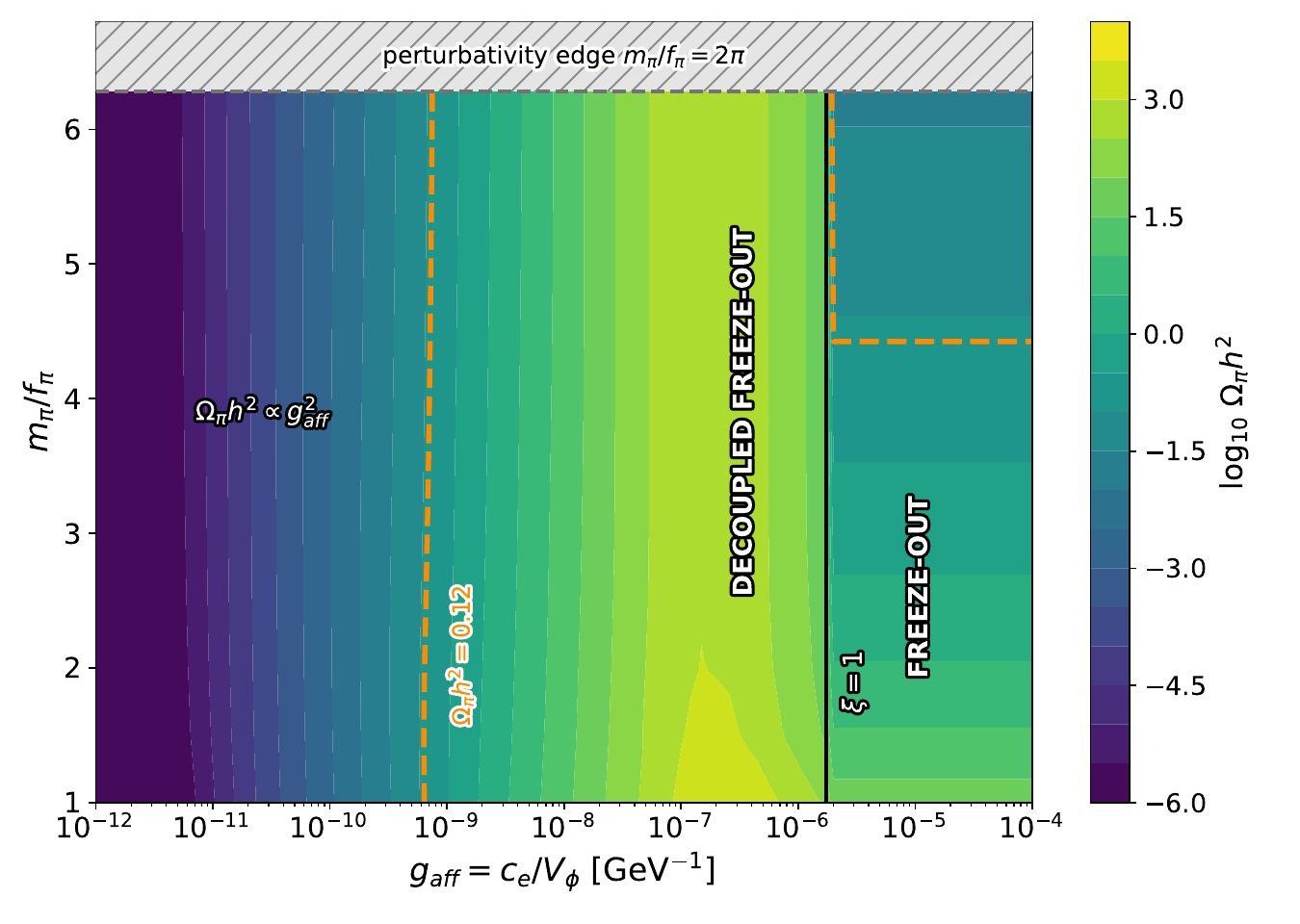}
  \caption{Phase diagram of the axion-portal SIMP in the plane of the connector
  coupling $g_{aff}=c_e/V_\phi$ and the hidden-sector strength $m_\pi/f_\pi$,
  drawn for the generic branch at $m_\pi=140\MeV$ and $m_a=1.25\,m_\pi=175\MeV$
  with the lepton charges of Eq.~\eqref{eq: lepto-charges}, so that the axis is
  the same coupling Sec.~\ref{sec: tied} quotes. Colour shows
  $\log_{10}\Omega_\pi h^2$ and the label in each region names the dominant
  process. The orange dashed line is the observed relic $\Omega_\pi h^2=0.12$. It
  runs vertically in the decoupled region, where the relic fixes $g_{aff}$ and
  hence $V_\phi$, and horizontally in the freeze-out region, where it fixes
  $f_\pi$ and reproduces the canonical SIMP miracle. The solid black line is the
  energy-transfer boundary $\xi=1$. The decoupled region is the output of the
  coupled solver of Sec.~\ref{sec: cannibal DFO} on a grid in $V_\phi$ and
  $m_\pi/f_\pi$, interpolated in logarithms and continued beyond the grid with
  the local slope, so the relic contour meets the benchmark of
  Table~\ref{tab: benchmark} by calculation rather than by construction. It does
  so to $5\%$, which is the grid resolution. The hatched band above
  $m_\pi/f_\pi=2\pi$ is where the chiral expansion is no longer perturbative. No freeze-in region appears: the internal $3\to2$
  rate exceeds the expansion rate throughout the plane, and the formal boundary
  $\Gamma_{3\to2}=H$ sits below $g_{aff}\sim10^{-22}\GeV^{-1}$ across the range
  shown, far to the left of it. The colour map is discontinuous across the
  $\xi=1$ line, since the two branches are computed under different assumptions.
  The decoupled branch conserves the dark entropy separately, while the
  freeze-out branch shares a single bath with the SM, and the two are not matched
  at the boundary. Along the bottom of the plane, at $m_\pi/f_\pi\lesssim3$, the
  weaker $3\to2$ rate brings dark freeze-out forward to visible temperatures of
  order $10^2\MeV$, before the injection has finished, so the separation between
  heating and cannibalism that Sec.~\ref{sec: cannibal DFO} describes does not
  hold there. The coupled solver carries the detailed-balance terms and handles
  that region, but the analytic account applies only above it.}
  \label{fig: mesa}
\end{figure}

We solve Eqs.~\eqref{eq: boltz-npi}--\eqref{eq: boltz-T}, scanning the connector
coupling $g_{aff}$ and the hidden-sector strength $m_\pi/f_\pi$ at fixed $m_\pi$
and $m_a$. Plotting the final abundance over this plane gives the phase diagram of
Fig.~\ref{fig: mesa}, which is the analogue for our model of the ``mesa'' diagram
of Refs.~\cite{Chu:2011be,Bharucha:2022lty}. The illustrative point is
$m_\pi=140\MeV$ and $m_a=1.25\,m_\pi$ with the lepton charges of
Eq.~\eqref{eq: lepto-charges}, so that the ALP is heavy enough for the inverse
decay $f\bar f\to a$ to be open and the horizontal axis is the same coupling
Sec.~\ref{sec: tied} quotes. Four production regimes appear in principle as the
portal is made stronger, and Table~\ref{tab: regimes} collects all four. Only two
of them fall inside the range the figure covers, since the internal $3\to2$ rate
is strong enough to push the threshold for chemical equilibrium within the dark
sector down to $g_{aff}\lesssim10^{-22}\GeV^{-1}$, far below the plotted range,
while connector freeze-out requires $f_a\lesssim f_\pi$ and plays no role for a
well-ordered spectrum. Across the couplings of Fig.~\ref{fig: mesa} the dark
matter is therefore made either by decoupled freeze-out or by ordinary
freeze-out. We first describe the physics of each of the four regions, since the
taxonomy is what organises the plane, and then give the conditions on the
couplings that separate them.

\subsection*{The freeze-in regime}
We begin with the region reached at the weakest portal. In the freeze-in regime,
at the smallest couplings, the dark pions never reach equilibrium. They are produced by inverse decays that populate the ALP, which then feeds $aa\to\pi\pi$. The $3\to2$ reaction never
activates because the density stays far below equilibrium. Since the pions are made
and not destroyed, their comoving yield is the production rate integrated over the
expansion history,
$Y_\pi(\infty) = \int_0^\infty \frac{\Gamma_{\rm prod}(T)}{s(T)\,H(T)\,T}\,dT$,
with $s(T)$ the visible entropy density. This yield is linear in the production cross
section $\braket{\sigma v}\propto g_{aff}^2$, giving the relic density
\begin{equation}
  \Omega_\pi h^2 = \frac{m_\pi s_0}{\rho_c/h^2}\, Y_\pi(\infty) \propto g_{aff}^2 ,
  \label{eq: freezein-scaling}
\end{equation}
where $s_0/(\rho_c/h^2)\simeq2.74\times10^8\GeV^{-1}$ is the entropy-to-critical-density
conversion factor. This quadratic growth with coupling is the hallmark of
freeze-in~\cite{Hall:2009bx}, causing the relic contour to run vertically in this
regime.
At the very smallest couplings the direct channel $f\bar f\to\pi\pi$ is subdominant
to the sequential one, in which the portal first makes ALPs through $f\bar f\to aa$
and the pions follow through $aa\to\pi\pi$, with the ALPs and pions never reaching a
common equilibrium. This regime does not appear in Fig.~\ref{fig: mesa}. The
internal $3\to2$ rate is strong enough that chemical equilibrium within the dark
sector is maintained at every coupling the figure covers, and the boundary
$\Gamma_{3\to2}=H$ lies below $g_{aff}\sim10^{-22}\GeV^{-1}$ throughout, off the
scale to the left.

\subsection*{The decoupled freeze-out regime}

A stronger portal reaches the decoupled freeze-out (DFO) regime, where the portal
populates and heats the hidden sector to a temperature $T^\prime=\xi T<T$ but never
brings it into equilibrium with the SM. The hidden sector
self-thermalises through the $3\to2$ reaction and through pion self-scattering, and
it then undergoes a cannibal $3\to2$ freeze-out at its own temperature.

The relic abundance is $\Omega_\pi h^2 = [m_\pi s_0/(\rho_c/h^2)]\,Y_\pi$, with $Y_\pi=n_\pi/s$ the
comoving pion yield and $s$ the visible entropy density. Since the two sectors are
decoupled it is convenient to refer the pion number to the dark entropy
$s^\prime$ by writing $Y_\pi=n_\pi/s=(n_\pi/s^\prime)(s^\prime/s)$. At freeze-out
the pions are non-relativistic and carry essentially all of the dark entropy,
$s^\prime\simeq n_\pi\,x^\prime_f$ with $x^\prime_f=m_\pi/T^\prime_f$, so that
$n_\pi/s^\prime\simeq1/x^\prime_f$. Using $m_\pi/x^\prime_f=T^\prime_f$ then gives
the decoupled relic abundance
\begin{equation}
  \Omega_\pi h^2 = \frac{s_0}{\rho_c/h^2}\,\frac{s^\prime}{s}\, T^\prime_f ,
  \label{eq: dfo-relic}
\end{equation}
where $T^\prime_f$ is the dark freeze-out temperature. The entropy ratio
$s^\prime/s$ is what the portal sets, and it is not conserved while the portal is
delivering energy. Section~\ref{sec: cannibal DFO} follows it through the
injection.

The freeze-out temperature is set by the $3\to2$ rate falling below Hubble,
\begin{equation}
  \braket{\sigma v^2}_{3\to2}\,n_\pi^2(T^\prime_f) = H .
  \label{eq: 3to2-fo}
\end{equation}
Since the equilibrium density is Boltzmann suppressed, $n_\pi\propto e^{-x^\prime_f}$,
this condition fixes the freeze-out point at
$x^\prime_f\simeq\tfrac12\ln(\braket{\sigma v^2}_{3\to2}/H)$ up to powers of
$x^\prime_f$. The exponential is steep enough that a change in the $3\to2$ coupling
is absorbed by a small shift in $x^\prime_f$, so the coupling enters $T^\prime_f$,
and through Eq.~\eqref{eq: dfo-relic} the relic, only logarithmically.

The entropy ratio itself is fixed by the portal. The energy the portal leaks into
the hidden sector is dominated by the inverse decay $f\bar f\to a$, a $2\to1$
reaction carrying one vertex on each side of the rate, so
$\rho^\prime/\rho\propto g_{aff}^2$. That energy arrives while the dark bath is
already non-relativistic, and for such a bath in chemical equilibrium the entropy
gained is linear in the energy delivered, so $s^\prime/s\propto g_{aff}^2$ as
well. We compute the energy transfer in Sec.~\ref{sec: relocated}, from the
collision integral given in Appendix~\ref{app: collision}, and follow the entropy
through the injection in Sec.~\ref{sec: cannibal DFO}.

Combining $s^\prime/s = (s^\prime/s)_0\,(g_{aff}/g_0)^2$ with
Eq.~\eqref{eq: dfo-relic} gives
\begin{equation}
  \Omega_\pi h^2 = \frac{s_0}{\rho_c/h^2}\,
    \left(\frac{s^\prime}{s}\right)_{\!0}
    \left(\frac{g_{aff}}{g_0}\right)^{2} T^\prime_f ,
  \label{eq: omega-gaff2}
\end{equation}
quadratic in the portal coupling up to the logarithm carried by $x^\prime_f$, and
nearly independent of $m_\pi/f_\pi$. Solving the coupled system we find the local
exponent $d\ln\Omega_\pi h^2/d\ln g_{aff}$ between $1.85$ and $1.92$ over three
decades in $V_\phi$, the shortfall below two coming from the drift of
$x^\prime_f$. As Fig.~\ref{fig: mesa} shows, the
observed abundance $\Omega_\pi h^2=0.12$ meets this region along a vertical line, so
in the decoupled regime the relic determines the ALP--SM coupling and
leaves $f_\pi$ free. Since $g_{aff}=c_f/V_\phi$ carries the flavon scale and
nothing else, this is already a determination of $V_\phi$, and
Sec.~\ref{sec: tied} shows that the surviving realisation keeps it that way.

\subsection*{The freeze-out and strong freeze-out regimes}

At stronger portals still the hidden sector shares the SM temperature,
$T^\prime=T$, and the relic is set by the ordinary $3\to2$ freeze-out of the SIMP
mechanism. It depends on $m_\pi/f_\pi$ and not on the portal, so the relic contour
runs horizontally and the observed abundance selects a value of $f_\pi$. This is the
standard SIMP miracle, which fixes the self-interaction cross section
$\sigma_{\rm self}/m_\pi$. However, this regime requires an ALP--SM
coupling in the range already probed by beam-dump and rare-decay searches for a
sub-GeV electrophilic ALP, as discussed in Sec.~\ref{sec: pheno}. At the strongest
couplings lies freeze-out driven by the connector, where $\pi\pi\to f\bar f$ sets
the relic as for a heavy weakly interacting particle. That regime requires
$f_a\lesssim f_\pi$ and plays no role for a well-ordered spectrum.

The central feature of Fig.~\ref{fig: mesa} is the shape of the relic curve, which
is vertical in the decoupled regime and horizontal in the freeze-out regime. In the
diagram of Ref.~\cite{Bharucha:2022lty} the relic curve is horizontal in the
decoupled regime, because the number-changing reaction there is a portal-mediated
$2\to2$ annihilation and the relic tracks the hidden coupling. In our model the
number-changer is the WZW $3\to2$, which is independent of the portal, so the relic
tracks $g_{aff}$ instead and the curve is rotated by ninety degrees. This rotation
is the geometric statement of the relocated miracle. In the freeze-out regime the
relic fixes $f_\pi$ and hence the self-interaction cross section, while in the
decoupled regime, where the observed relic actually sits for a feebly coupled and
viable portal, it fixes $g_{aff}$ instead.

The four cosmological regimes, along with their
governing thermal relations, dominant number-changing mechanisms, and relic scalings,
are summarised in Table~\ref{tab: regimes}.

\begin{table}[t]
  \centering
  \renewcommand{\arraystretch}{1.35}
  \setlength{\tabcolsep}{4pt}
  \begin{tabular}{|>{\raggedright\arraybackslash}p{2.2cm}|>{\centering\arraybackslash}p{3.4cm}|>{\centering\arraybackslash}p{2.4cm}|>{\centering\arraybackslash}p{2.6cm}|>{\centering\arraybackslash}p{2.5cm}|}
    \hline
    {\bf Regime} & {\bf Portal coupling $g_{aff}$} & {\bf Bath relation} & {\bf Number-changing process} & {\bf Relic scaling $\Omega_\pi h^2$} \\
    \hline
    Sequential Freeze-in & far below the relic line & $T^\prime \ll T$ (non-thermal) & $f\bar f\to aa\to\pi\pi$ & $\propto g_{aff}^2$ \\
    Direct Freeze-in     & far below the relic line & $T^\prime \ll T$ (non-thermal) & $f\bar f\to\pi\pi$       & $\propto g_{aff}^2$ \\
    Decoupled Freeze-out (DFO) & up to $1.7\times10^{-6}\GeV^{-1}$ & $T^\prime < T$ & Cannibal $3\to2$ & $\propto (s^\prime/s)\,T^\prime_f \propto g_{aff}^{2}$ \\
    Standard Freeze-out  & $1.7\times10^{-6}$--$29\GeV^{-1}$ & $T^\prime = T$ (contact) & SIMP $3\to2$     & $\propto f_\pi^5 / m_\pi^{7/2}$ \\
    Connector Freeze-out & 
   $\gtrsim29\GeV^{-1}$ & $T^\prime = T$ & $\pi\pi\to f\bar f$ (annihilation) & $\propto 1/g_{aff}^2$ \\
    \hline
  \end{tabular}
  \caption{Summary of the dark matter production regimes across the axion-portal parameter space of Fig.~\ref{fig: mesa}. The freeze-in boundary is quoted only as an ordering, since locating it requires the freeze-in yield rather than the equilibrium density used to draw it, and it lies far below the observed relic line in any case. In the decoupled freeze-out regime, cannibalism sets the relic through the entropy the portal injects into the dark bath, relocating the standard SIMP miracle from $f_\pi$ to the portal coupling $g_{aff}$. The quadratic scaling there carries a logarithmic correction through $x^\prime_f$, which brings the measured exponent to $1.85$--$1.92$. The two rows that run through $\pi\pi\to f\bar f$, direct freeze-in and connector freeze-out, require a trilinear $a\pi\pi$ vertex and are absent from the Hermitian portal of Sec.~\ref{sec: model}; they are listed because the taxonomy is general.}
  \label{tab: regimes}
\end{table}

\subsection*{The freeze-in/decoupled freeze-out boundary}
These four regions are separated by three boundaries, each fixed by a competition of
rates. The first separates freeze-in from decoupled freeze-out. It is the
condition that the hidden sector reaches internal chemical equilibrium through the
$3\to2$ reaction,
\begin{equation}
  \braket{\sigma v^2}_{3\to2}\, n_\pi^2 = H
  \qquad\text{at}\qquad T^\prime\simeq m_\pi ,
  \label{eq: fi-dfo}
\end{equation}
where the condition estimates the onset of dark-sector internal chemical
equilibrium before cannibal freeze-out. Below this boundary the injected pion
yield is insufficient for the three-to-two process to maintain chemical
equilibrium against the Hubble expansion, whereas above it the dark sector
thermalises internally. As the three-to-two interaction strength $m_\pi/f_\pi$
increases, a smaller injected density suffices to reach internal equilibrium, so
the boundary moves towards smaller couplings. For the parameters of
Fig.~\ref{fig: mesa} it sits below $g_{aff}\sim10^{-22}\GeV^{-1}$ across the whole
range of $m_\pi/f_\pi$, which is why no freeze-in region is visible there. We
quote its position only as an ordering, since Eq.~\eqref{eq: fi-dfo} tests
chemical equilibrium against the equilibrium density, whereas the freeze-in yield
is what actually populates the sector below the boundary. This does not affect the
conclusions, because the observed relic lies many orders of magnitude above it.

\subsection*{The decoupled/ordinary freeze-out boundary}
The second boundary separates decoupled freeze-out from ordinary freeze-out. It is
the condition that the two sectors reach a common temperature at the epoch of
injection, $T^\prime/T\to1$. At this benchmark the energy that the portal
delivers to the hidden sector is dominated by the inverse decay $e^+e^-\to a$,
whose integrand peaks at $T=2m_a/9$, so the injected density scales as
$\rho^\prime/\rho\propto g_{aff}^2$. At the benchmark the boundary is reached at
$g_{aff}\simeq1.7\times10^{-6}\GeV^{-1}$. While the injection is running this is
a ceiling rather than a crossing, since the dark sector is populated only by the
energy that leaks from the SM, so it can at most reach the SM temperature, and a
hotter dark sector would require a preferential reheating of the hidden sector
that we do not assume. It is not a ceiling afterwards. The cannibal phase cools
the dark bath only logarithmically while the visible one cools with the scale
factor, so $T^\prime/T$ rises through the cannibal phase and passes unity before
freeze-out. At the benchmark it is $1.05$ where
$\Gamma_{3\to2}=H$, at $T=6.1\MeV$, and continues to rise to a maximum of $1.45$
at $T=3.6\MeV$, where $\Gamma_{3\to2}/H=6\times10^{-2}$. The maximum sits after
the nominal crossing rather than at it, because the residual $3\to2$ conversion
of rest mass into kinetic energy still slows the dark cooling for about half a
decade past $\Gamma_{3\to2}=H$. Only once it is negligible do the pions
free-stream with $T^\prime\propto a^{-2}$ and $T^\prime/T$ fall. This costs
nothing energetically, since the dark sector holds
$\rho^\prime/\rho_{\rm SM}=1.6\times10^{-8}$ at $T=40\MeV$ and
$6.8\times10^{-7}$ at $T=1\MeV$, and it is a further reason to treat
$T^\prime/T$ as an output rather than as a parameter. This divide is a band rather than a sharp line. The sectors reach a common temperature
by energy transfer at the $\xi=1$ line, but the elastic scattering that must
maintain that common temperature through freeze-out, $\Gamma_{\pi a\to\pi a}$ and
$\Gamma_{\pi f\to\pi f}$, only exceeds $H$ at a much stronger coupling,
$g_{aff}\simeq0.05\GeV^{-1}$ for this benchmark. Between the two the hidden sector
is heated early but drifts to a lower temperature before freeze-out, and we treat
this band explicitly in Sec.~\ref{sec: consistency}.

\subsection*{The ordinary/connector freeze-out boundary}
The third boundary separates freeze-out driven by the hidden $3\to2$ from
freeze-out driven by the connector. The reaction $\pi\pi\to f\bar f$ overtakes the
$3\to2$ process only when $f_a\lesssim f_\pi$, that is when $g_{aff}\gtrsim1/f_\pi$, which at the benchmark is $g_{aff}\gtrsim29\,\GeV^{-1}$, far to the right of the plot and far above any
allowed coupling. In practice the $3\to2$ process always wins for $f_a\gg f_\pi$,
which is the only sensible ordering of the two decay constants.

\begin{figure}[t]
  \centering
  \includegraphics[width=0.9\columnwidth]{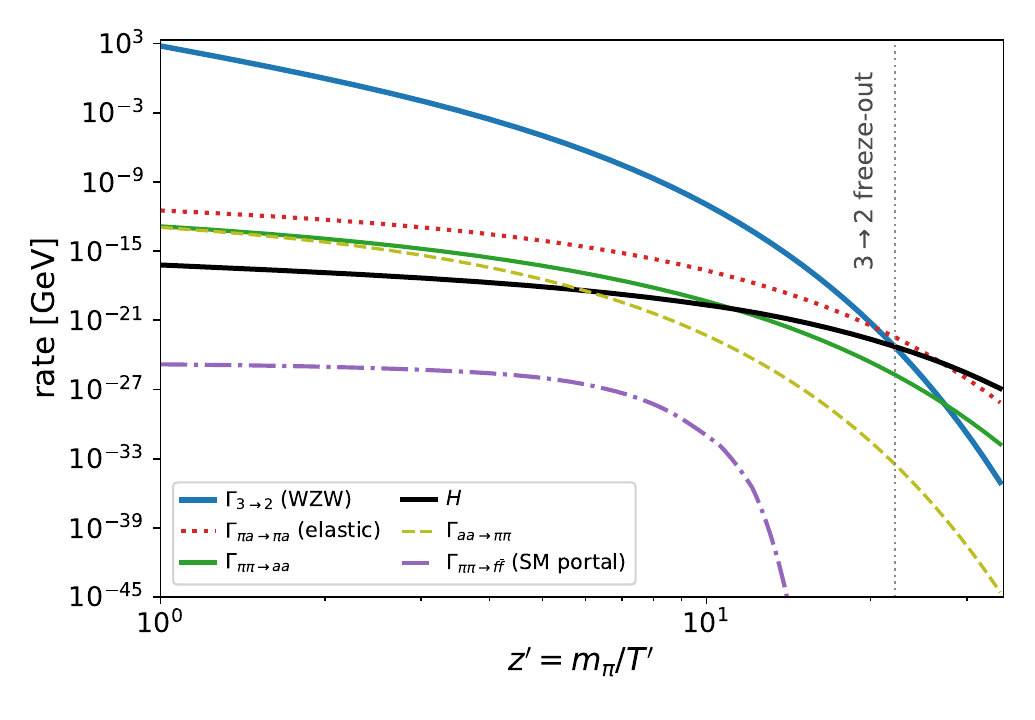}
  \caption{Interaction rates in the dark sector at a decoupled-freeze-out benchmark
  ($m_\pi=100\MeV$, $f_a=3\GeV$, $m_a=589\keV$, $\xi=0.014$, and $V_\phi=f_a$),
  as functions of
  $z^\prime=m_\pi/T^\prime$. The WZW $3\to2$ rate dominates the number-changing over
  $\pi\pi\to aa$, the elastic $\pi a\to\pi a$ rate keeps the dark sector at a single
  temperature, and the portal $\pi\pi\to f\bar f$ lies far below Hubble. Freeze-out
  of the $3\to2$ process is at $z^\prime\simeq22$. The figure is drawn for the
  generic model, in order to illustrate the rate hierarchy, and the ALP mass shown
  is the condensate-tied value. Section~\ref{sec: nogo} excludes that tie, and in
  the benchmark that survives the ALP sits at $m_a=1.25\,m_\pi$, where
  $\pi\pi\to aa$ is Boltzmann suppressed and $\pi\pi\to f\bar f$ vanishes with the
  trilinear.}
  \label{fig: wzw rates}
\end{figure}

Figure~\ref{fig: wzw rates} makes the dominance of the $3\to2$ reaction explicit at
a decoupled benchmark. The WZW $3\to2$ rate exceeds the internal conversion
$\pi\pi\to aa$ by about two orders at freeze-out, so it is the WZW term and not the portal that depletes the pion number.
The elastic $\pi a\to\pi a$ rate stays far above Hubble throughout, which holds the
pions and ALPs at a common temperature, while the portal $\pi\pi\to f\bar f$ runs
many orders below Hubble, so the sector is decoupled from the SM.
Switching off the WZW term would remove the $3\to2$ channel altogether, and the
relic would instead be set by the $2\to2$ annihilation $\pi\pi\to aa$, which is a
different mechanism and not the SIMP miracle. The WZW term is therefore
indispensable to the scenario.

%% file: sec_consistency.tex

\section{The two branches and the kinetic-contact boundary}
\label{sec: consistency}

\begin{figure}[t]
  \centering
  \includegraphics[width=\columnwidth]{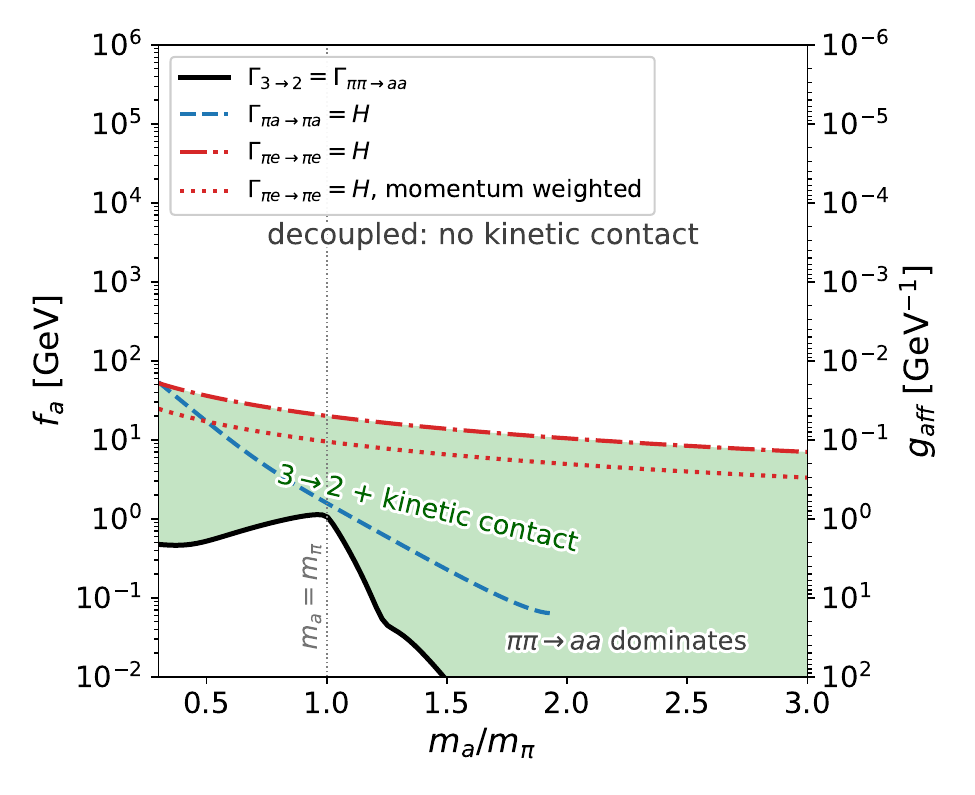}
  \caption{Boundaries in the plane of $m_a/m_\pi$ and the ALP decay constant
  $f_a$, for $m_\pi=100\MeV$, $f_\pi=20\MeV$, $c_f=1$ and dark freeze-out
  $x_f^\prime=20$, with the right-hand axis giving the connector coupling
  $g_{aff}=c_f/f_a$, again under the condition $V_\phi=f_a$. Above the
  solid black curve the $3\to2$ process dominates
  $\pi\pi\to aa$. Below the coloured curves the elastic channels $\pi a\to\pi a$
  and off-shell $\pi f\to\pi f$ hold kinetic contact. The $\pi a\to\pi a$ curve is shown
  only while the thermal ALP population is numerically resolved, since it is
  Boltzmann suppressed for $m_a>m_\pi$. The off-shell $\pi f\to\pi f$ boundary
is drawn for a portal carrying a trilinear $a\pi\pi$ vertex. The Hermitian portal
  of Eq.~\eqref{eq: portal-op} carries none, in either mass case, so this channel
  is closed and kinetic contact rests entirely on the thermal ALP population, which
  only strengthens the conclusion that the sector decouples.
  The shaded band is the region where the
  $3\to2$ process sets the relic and kinetic contact survives. It exists for all
  $m_a/m_\pi$ and widens for $m_a>m_\pi$, but it lies entirely at
  $f_a\lesssim20\GeV$, that is $g_{aff}\gtrsim0.05\GeV^{-1}$.}
  \label{fig: regimemap}
\end{figure}

The phase diagram of Sec.~\ref{sec: regimes} shows that the observed relic sits in
the decoupled branch for a feeble portal and in the freeze-out branch for a strong
one. What decides between them is whether the dark sector holds kinetic contact
with the SM through freeze-out, and in this section we locate that
divide. We fix $m_\pi$, $f_\pi$ and $c_f$ and work in the plane of the ALP mass
ratio $m_a/m_\pi$ and the portal coupling, which is the generic case of
Sec.~\ref{sec: model} in which a bare mass term is present. The result is
Fig.~\ref{fig: regimemap}.

The first boundary is the condition that the $3\to2$ process dominates the
hidden-sector annihilation. The two reactions have a different number of initial
particles, so the rate per pion carries two powers of the density for the $3\to2$
process and one power for $\pi\pi\to aa$, and the two are equal when
\begin{equation}
  \braket{\sigma v^2}_{3\to2}\, n_\pi^2 = \braket{\sigma v}_{\pi\pi\to aa}\, n_\pi ,
  \label{eq: 3to2-vs-2to2}
\end{equation}
where both sides are evaluated at the freeze-out temperature
$T^\prime\simeq m_\pi/20$. The right-hand side scales as $1/f_a^4$, so the $3\to2$
process wins for $f_a\gtrsim f_\pi$. At the benchmark this is $f_a\gtrsim1\GeV$
for $m_a=m_\pi$, and it falls steeply for $m_a>m_\pi$. The claim sometimes
made, that $m_a>m_\pi$ is required and that $\pi\pi\to aa$ overwhelms the $3\to2$
process as soon as it opens, does not hold. For $m_a>m_\pi$ the channel
$\pi\pi\to aa$ is not closed but kinematically threshold-suppressed and Boltzmann suppressed. At
$m_a=1.1\,m_\pi$ the suppression at freeze-out is only $e^{-0.1x}\sim e^{-2}$, so
the ratio of the two rates changes by little more than an order of magnitude
across threshold, and it reaches four to six orders only near
$m_a\simeq1.3$--$1.5\,m_\pi$.

The second and third boundaries test whether kinetic contact with the SM
survives. Contact does not require the on-shell ALP, since it is carried by
two elastic channels. The first is $\pi a\to\pi a$ on the thermal ALP population,
whose rate scales with $n_a$, which for $m_a<m_\pi$ is enhanced relative to $n_\pi$
by
\begin{equation}
  \frac{n_a}{n_\pi}\sim e^{(m_\pi-m_a)/T^\prime} ,
  \label{eq: na-over-npi}
\end{equation}
where the exponent is a large number at freeze-out. This channel shuts off once
$m_a>m_\pi$. The second is the off-shell $\pi f\to\pi f$ scattering on
relativistic SM fermions, dominated by electrons. It exchanges an ALP in the $t$ channel, so its squared amplitude carries $1/(t-m_a^2)^2$ from the propagator, $1/f_a^4$ from its two vertices, and $\tan^2\!\bar\theta$ from the trilinear. However, it carries no ALP-mass Boltzmann suppression, because the ALP is
never on shell, and it is fed by the large fermion number density. It therefore
persists for $m_a>m_\pi$, where the first channel has closed. It requires the
trilinear $a\pi\pi$ vertex, which Sec.~\ref{sec: tied} shows the Hermitian portal
does not generate in either mass case, so this channel is absent altogether
and contact is lost more easily than the boundaries drawn here suggest. We keep it
in the generic map, where the ALP mass and the portal are independent, and obtain
its squared amplitude by crossing the $\pi\pi\to f\bar f$ amplitude of
Appendix~\ref{app: xsec},
\begin{equation}
  |\mathcal{M}|^2_{\pi f\to\pi f}
      = \tan^2\!\bar\theta\;
        \frac{c_f^2\, m_f^2\, m_\pi^4\,(-t)}{2\,f_a^4\,(t-m_a^2)^2},
  \label{eq: pif}
\end{equation}
where $t$ is the usual Mandelstam variable. Each channel defines a decoupling
boundary through $\Gamma=H$ at freeze-out, and since both rates scale as $1/f_a^4$
contact survives only below a critical $f_a$. At the benchmark this is
$f_a\lesssim20\GeV$, with a mild dependence on $m_a/m_\pi$ through the off-shell
channel.

The two families of curves do not close. There is a band in which the $3\to2$
process sets the relic and kinetic contact survives, and this is the freeze-out
branch of Sec.~\ref{sec: regimes}. It is narrow for $m_a<m_\pi$, where contact
rests on the enhanced $\pi a\to\pi a$ channel, and it widens for $m_a>m_\pi$, where
the off-shell $\pi f\to\pi f$ channel takes over. The band exists for any ALP mass.
What it does not do is move to weak coupling, since it lies entirely at
$f_a\lesssim20\GeV$, which corresponds to an ALP--SM coupling
$g_{aff}=c_f/f_a\gtrsim0.05\GeV^{-1}$. For a sub-GeV electrophilic ALP this is in
the range already probed by beam-dump and rare-decay searches, as discussed in
Sec.~\ref{sec: pheno}.

The divide between the two branches is not a single line. The energy transfer
brings the two sectors to a common temperature at $\xi=1$, but the elastic
scattering that must maintain that temperature through freeze-out only exceeds $H$
at a stronger coupling. Between the two the hidden sector is heated early and then
drifts to a lower temperature before freeze-out. We treat this transition band in
the numerical solution and quote the two edges where they differ.

It is worth locating the two versions of the minimal model in this plane. The
tied-mass constraint of Eq.~\eqref{eq: tied-min} is the hyperbola
$m_a/m_\pi=\sqrt{N_f}f_\pi/2f_a$, which for $f_\pi=20\MeV$ gives
$m_a/m_\pi\simeq14\MeV/f_a$. Even at the upper edge of the contact band,
$f_a\simeq20\GeV$, this is $m_a/m_\pi\sim7\times10^{-4}$, so the tied model lies
far to the left of Fig.~\ref{fig: regimemap} everywhere, and contact in that
corner would rest entirely on the exponentially enhanced $\pi a\to\pi a$ channel.
Section~\ref{sec: tied} shows that this corner is disfavoured for a different
reason, and that the realisation which survives sits instead near
$m_a/m_\pi\simeq1.25$,
just to the right of the threshold in this figure, where the same $\pi\pi\to aa$
channel is Boltzmann suppressed.

The conclusion of this section is unaffected by which of the two applies, and is
therefore structural. Decoupling is not forced by the ALP mass. It is the generic
outcome once the portal is weak enough to evade the laboratory bounds, and it then
holds for any $m_a/m_\pi$. The remainder of the paper works in this decoupled
branch, where the observed relic lies.

%% file: sec_cannibal.tex

\section{Cannibal freeze-out in the decoupled sector}
\label{sec: cannibal DFO}

In the decoupled branch the portal is too weak to hold the dark sector at the
SM temperature. The dark sector then evolves at its own temperature
$T^\prime$ and undergoes the $3\to2$ freeze-out as a cannibal gas. This
section derives the relic abundance in that regime and isolates what it depends
on.

Two internal reactions keep the dark sector in equilibrium with itself. The
number-changing $3\to2$ reaction sets chemical equilibrium, and the elastic pion
self-scattering sets kinetic equilibrium. The self-scattering rate is
$\Gamma_{\rm self}=n_\pi\,\sigma_{\rm self}v$ with
$\sigma_{\rm self}\simeq m_\pi^2/32\pi f_\pi^4$~\cite{Hochberg:2014kqa}, where
$\sigma_{\rm self}$ is the same cross section that the SIMP mechanism requires to
be strong. At freeze-out it exceeds the Hubble rate by some nine orders of
magnitude, so the single temperature $T^\prime$ is well defined for the pions
throughout freeze-out. Kinetic contact with the SM plays no role here, since it
has already been lost.

The ALP is not a spectator while the portal is injecting. The energy the portal
delivers arrives in ALPs, and it reaches the pions through the elastic scattering
$\pi a\to\pi a$ and the conversion $aa\to\pi\pi$, both of which follow from the
contact coupling $g_{aa\pi\pi}=m_\pi^2/4f_a^2$ of
Eq.~\eqref{eq: portal-cos}. At the benchmark, and at the temperature where the
injection peaks, and at $f_a=3\GeV$, the elastic rate per ALP exceeds the Hubble
rate by $6\times10^{2}$ and the conversion rate by $10$, so the ALP
and the pions do share a temperature while the energy is being delivered. Both rates fall
as $1/f_a^4$, so requiring that they survive places a ceiling on the dark scale,
which we compute in Sec.~\ref{sec: tied}. By dark freeze-out, an order of
magnitude lower in temperature, both have fallen below $H$, and the ALP no longer
feeds back on the pion number. Because the ALP carries at most a few degrees of
freedom at a temperature well below $T$, it contributes negligibly to the
radiation density, as quantified in Sec.~\ref{sec: pheno}.

While the $3\to2$ reaction is faster than expansion it holds the pions in
chemical equilibrium at $T^\prime$ with vanishing chemical potential, and at the
benchmark $\Gamma_{3\to2}/H\simeq3.5\times10^{3}$ where the injection peaks. A
$3\to2$ reaction in equilibrium is reversible and generates no entropy of its
own, so the comoving dark entropy is conserved once the portal has stopped
delivering energy. It is not conserved while the portal operates. With $\mu=0$
the first law gives $T^\prime dS^\prime=dE^\prime$, and integrating downwards in
the visible temperature the comoving entropy ratio obeys
\begin{equation}
  \frac{d}{dT}\left(\frac{s^\prime}{s}\right)
  = -\,\frac{C_E(T)}{H(T)\,T\,s(T)\,T^\prime(T)} ,
  \label{eq: entropy-injection}
\end{equation}
where $C_E$ is the energy-transfer collision integral of
Sec.~\ref{sec: regimes}, $s$ is the visible entropy density and $s^\prime$ that
of the dark bath.

The dark bath is not relativistic when this energy arrives, and that is what
fixes the scaling. At the benchmark $T^\prime$ never rises above about
$10\MeV$, so $x^\prime=m_\pi/T^\prime$ exceeds $14$ throughout the injection
window, and at the peak of the injection integrand, $T=2m_a/9=39\MeV$, it is
$x^\prime=16.5$. For a non-relativistic bath Eq.~\eqref{eq: entropy-injection}
makes the entropy gained linear in the energy delivered, rather than the
three-quarter power a relativistic bath would give.

The abundance is $\Omega_\pi h^2\propto m_\pi Y_\pi$ with $Y_\pi=n_\pi/s$, and
since the two sectors are decoupled it is convenient to refer the pion number to
the dark bath, $Y_\pi=(n_\pi/s^\prime)(s^\prime/s)$. Freeze-out occurs when the
$3\to2$ rate falls below the Hubble rate,
\begin{equation}
  \braket{\sigma v^2}_{3\to2}\, n_\pi^2(T^\prime_f) = H(T_f) ,
  \label{eq: fo-condition}
\end{equation}
where $T^\prime_f$ and $T_f$ are the dark and visible temperatures at that
moment. At that point the non-relativistic pions carry essentially all of the
dark entropy, $s^\prime\simeq n_\pi x^\prime_f$, so that
$n_\pi/s^\prime\simeq1/x^\prime_f$, and using $m_\pi/x^\prime_f=T^\prime_f$ gives
\begin{equation}
  \Omega_\pi h^2 = \frac{s_0}{\rho_c/h^2}\,\frac{s^\prime}{s}\,T^\prime_f ,
  \label{eq: relic-entropy}
\end{equation}
where $s_0/(\rho_c/h^2)\simeq2.74\times10^8\GeV^{-1}$ is the
entropy-to-critical-density conversion factor.

Two powers of the portal coupling enter through $C_E$, since the inverse decay
$f\bar f\to a$ that dominates the injection is a $2\to1$ reaction carrying one
vertex on each side of the rate. Since $T^\prime$ is nearly constant across the
injection window, Eq.~\eqref{eq: entropy-injection} then gives
$s^\prime/s\propto g_{aff}^2$, while the freeze-out condition
Eq.~\eqref{eq: fo-condition} with
$\braket{\sigma v^2}_{3\to2}\propto m_\pi^5/f_\pi^{10}$ fixes $x^\prime_f$, and
hence $T^\prime_f$, only inside a logarithm. The relic abundance therefore obeys
\begin{equation}
  \Omega_\pi h^2 \;\propto\; \frac{g_{aff}^2}{x^\prime_f} ,
  \label{eq: relic-scaling}
\end{equation}
quadratic in the portal coupling up to that logarithm, and independent of
$m_\pi$ and $f_\pi$ except through the same logarithm. Solving the coupled
system numerically we find
$d\ln\Omega_\pi h^2/d\ln g_{aff}=1.85$ to $1.92$ across
$V_\phi=3\times10^9$ to $10^{11}\GeV$, the shortfall below two coming from the
drift of $x^\prime_f$, and the exponent agrees to one per cent at
$m_\pi/f_\pi=2$, $4$ and $6$.

Had the injection finished while the dark bath was still relativistic, the
entropy would have followed $s^\prime\propto T^{\prime3}\propto\rho^{\prime3/4}$,
the temperature ratio $\xi=(\rho^\prime/\rho)^{1/4}$ would have been a constant of
the problem, and the same counting would have given
$\Omega_\pi h^2\propto\xi^3\propto g_{aff}^{3/2}$. That is not the case here, and
$T^\prime/T$ is not a constant either. It rises through the cannibal phase, since
$T^\prime$ falls only logarithmically while $T$ falls with the scale factor, and
at the benchmark it passes through unity before freeze-out. We therefore quote
$\xi$ only as the value of $T^\prime/T$ at the start of the integration,
$T=3m_\pi$, and use $s^\prime/s$ wherever a scaling is needed.

Two consequences follow. First, the relic abundance no longer determines $f_\pi$.
In the standard SIMP mechanism matching the observed relic abundance eliminates
$f_\pi$ and fixes the self-interaction cross section $\sigma_{\rm self}/m_\pi$,
whereas here it fixes the injected entropy instead and leaves $f_\pi$ free. The
chain $\Omega h^2\to f_\pi\to\sigma_{\rm self}/m_\pi$ is therefore broken. Second,
the mass window opens. With $\sigma_{\rm self}/m_\pi$ no longer tied to the relic,
the dark pion mass is bounded only by the bullet cluster from below and by
perturbativity from above, and it can rise towards the GeV scale, where the chiral
expansion is better controlled. This recovers, from an explicit portal, the
larger-mass self-interacting window anticipated in
Ref.~\cite{Heikinheimo:2018esa}, where it followed from treating $\xi$ as a free
input. Here it follows from an injection that the portal fixes, which is the
subject of Sec.~\ref{sec: relocated}.

%% file: sec_relocated.tex

\section{Relocating the miracle}
\label{sec: relocated}

Section~\ref{sec: cannibal DFO} showed that in the decoupled branch the relic
fixes the entropy the portal injects into the dark sector and not the decay
constant. This looks like a loss of predictivity. However, it is not, because
the injected entropy is itself set by the portal. In this section we compute it
as a function of the ALP--SM coupling and turn matching the observed relic
abundance into a prediction for that coupling.

The energy that the portal leaks from the SM into the dark sector follows from
the energy-transfer collision integral of Sec.~\ref{sec: regimes}, integrated
from the reheating temperature down,
\begin{equation}
  \frac{\rho^\prime}{\rho}(T)
  = \int_T^{T_{\rm RH}}\!\! d\tilde T\;
    \frac{C_E(\tilde T)}{H(\tilde T)\,\tilde T\,\rho(\tilde T)} ,
  \label{eq: rhop}
\end{equation}
where $C_E$ collects the inverse decays and the $f\bar f\to aa$ energy transfer,
and $T_{\rm RH}$ is the reheating temperature. What the relic responds to is not
this ratio but the entropy it deposits through
Eq.~\eqref{eq: entropy-injection}, and the temperature ratio $T^\prime/T$ is an
output of the coupled system rather than a parameter of it.

Which power of $g_{aff}$ appears in Eq.~\eqref{eq: rhop} depends on whether the
inverse decay $f\bar f\to a$ is kinematically open. When the ALP is heavy enough,
that is $m_a>2m_f$ for at least one fermion, the integral is dominated by the
inverse decay. That is a $2\to1$ reaction carrying one vertex on each side of the
rate, so
\begin{equation}
  \frac{\rho^\prime}{\rho}\propto g_{aff}^2 ,
  \label{eq: xi-scaling-open}
\end{equation}
and with the dark bath non-relativistic when the energy arrives,
Eq.~\eqref{eq: relic-scaling} gives $\Omega_\pi h^2\propto g_{aff}^2$ up to the
logarithm carried by $x^\prime_f$. When the inverse decay is closed, the
injection proceeds instead through the $2\to2$ transfer $f\bar f\to aa$, which
uses the ALP--fermion vertex twice and so carries four powers of the coupling,
\begin{equation}
  \frac{\rho^\prime}{\rho}\propto g_{aff}^4 .
  \label{eq: xi-scaling-closed}
\end{equation}
The relic scaling that follows from Eq.~\eqref{eq: xi-scaling-closed} depends on
whether the dark bath is still relativistic when that energy arrives, which is a
different question at a different ALP mass, and we do not need it, since
Sec.~\ref{sec: tied} excludes the branch in which the inverse decay is closed.
The transfer $f\bar f\to\pi\pi$ plays no role in either case, since it requires
the trilinear $a\pi\pi$ vertex that Sec.~\ref{sec: tied} shows the portal does
not generate.

At the benchmark of Sec.~\ref{sec: tied} the ALP sits at $m_a=175\MeV$, below the
dimuon threshold $2m_\mu=211\MeV$, so the only inverse decays open are
$e^+e^-\to a$ and $\gamma\gamma\to a$, the latter carrying $3.1\%$ of the total
width. The integrand of Eq.~\eqref{eq: rhop} then behaves as
$T^{-9/2}e^{-m_a/T}$ per logarithmic interval and peaks at $T=2m_a/9=39\MeV$,
with $86\%$ of the energy delivered below $3m_\pi$ and $40\%$ below $m_\pi$. The
injection is therefore infrared dominated, and we find
$\rho^\prime/\rho$ unchanged at the per cent level as $T_{\rm RH}$ runs from
$10^2$ to $10^6\GeV$. Since $\Omega_\pi h^2$ is close to quadratic in the
coupling, a shift in the normalisation of the injection moves $V_\phi$ by about
half as much in relative terms, so the flavon scale inherits none of the
reheating history.

Matching the observed relic abundance $\Omega_\pi h^2=0.12$ therefore selects a
value of $g_{aff}$, and the statement ``the relic fixes the injected entropy''
becomes ``the relic fixes $g_{aff}$''. This is the sense in which the miracle is
relocated. In the standard SIMP mechanism the relic predicts a self-interaction
cross section, which is accessible only through astrophysical systematics, while
here it predicts an ALP--SM coupling, which is the target of a definite programme
of laboratory and astrophysical probes. The injection carries the flavon scale
alone and not the dark scale, since neither the inverse decay nor
$f\bar f\to aa$ involves the portal to the pions, so the relic condition fixes
$V_\phi$ outright rather than a product of two scales. Solving the full system
with the leptonic charges of Sec.~\ref{sec: tied}, $V_\phi$ stays within $16\%$
of $1.15\times10^{10}\GeV$ across $m_\pi=100$ to $140\MeV$, $m_\pi/f_\pi=2$ to
$6$ and $f_a=1$ to $10\GeV$, which is what Eq.~\eqref{eq: relic-scaling} predicts.

Two caveats belong here. First, the absolute normalisation of $g_{aff}$ depends
on the full set of production channels. The estimate here uses inverse decays and
the tree-level $2\to2$ transfers, so the radiative and gauge-boson channels
discussed in Sec.~\ref{sec: model} could shift it. The scaling
Eq.~\eqref{eq: relic-scaling} is unaffected, because it follows from the coupling
counting and the equation of state of the dark bath rather than from the absolute
rate. Second, the prediction is a band and not a point, since $f_\pi$ enters
$x^\prime_f$ logarithmically, so a range of $f_\pi$ maps to a narrow range of
$g_{aff}$ at fixed relic abundance. The band is the $16\%$ quoted above, and it is
smaller than the $\mathcal{O}(1)$ uncertainty the coset normalisation of
Sec.~\ref{sec: pheno} already carries.

%% file: sec_tied.tex

\section{The minimal model}
\label{sec: tied}

The previous sections treated the ALP mass as an input. We now ask what the
minimal realisation gives, in which the ALP is the axiflavon of
Eq.~\eqref{eq: flavon} and its mass comes from the portal of
Eq.~\eqref{eq: portal-op}. We find that the answer turns on a single structural
feature of that portal, namely that it is even in the ALP field, and that this
feature divides the model into two cases with opposite fates. Throughout this
section we take $N_c=2$, $N_f=2$ and $m_\pi/f_\pi=4$.

\subsection{The minimal realisation}
\label{sec: nogo}

The operator of Eq.~\eqref{eq: portal-op} carries its Hermitian conjugate, and
the invariant it contains is real. The broken generators of the $SU(4)/Sp(4)$
coset are traceless and satisfy ${\rm Tr}[\pi^3]=0$, and the Cayley--Hamilton
theorem for a traceless $4\times4$ matrix then gives ${\rm Tr}[\pi^{2k+1}]=0$ for
every odd power, so ${\rm Tr}[J\Sigma]=-{\rm Tr}[e^{2i\pi/f_\pi}]$ is real. The
two exponentials therefore combine into a cosine,
\begin{equation}
  \mathcal{L}_{a\pi}
  = -\tfrac12 m_{\rm PQ}^2 a^2
    + m_Q\mu^3\,\cos\!\Big(\frac{a}{f_a}\Big)\,
      {\rm Tr}\Big[\cos\!\Big(\frac{2\pi}{f_\pi}\Big)\Big] ,
  \label{eq: portal-cos}
\end{equation}
which is even in $a$ and even in $\pi$ separately. Every vertex with an odd
number of ALP legs cancels against its conjugate, so the couplings $a\pi\pi$,
$a^3$ and $a^3\pi\pi$ that a single exponential would generate are absent. What
survives is the quartic contact term of Eq.~\eqref{eq:ALP-DM L}, and the dark
sector therefore couples to the ALP only in pairs.

This is not a technicality. The contact term carries no pole, so the conversion
$\pi\pi\to aa$ proceeds through a single diagram whose squared amplitude is
independent of the scattering angle, and the elastic channel $\pi a\to\pi a$
follows from it by crossing. Both scale as $1/f_a^4$, as before, but the
near-cancellation between contact and pion exchange that a trilinear coupling
would provide is gone, and we find that the conversion is some four orders of
magnitude faster than the trilinear estimate.

The ALP mass follows from the same operator together with any explicit breaking
above the condensate scale,
\begin{equation}
  m_a^2 = m_{\rm PQ}^2 + \frac{N_f\,m_\pi^2 f_\pi^2}{4 f_a^2} ,
  \label{eq: ma-two-pieces}
\end{equation}
and the two terms define the two cases. In the \emph{tied} case
$m_{\rm PQ}=0$ and the flavour condensate alone sets the mass, which for
$f_a\gg f_\pi$ places the ALP far below the dark pion. In the \emph{generic}
case the explicit term dominates and $m_a$ is a free parameter. We take the two
in turn, and the tied one closes first.

With $m_{\rm PQ}=0$ the tie reads $m_a=\sqrt{N_f}\,m_\pi f_\pi/2f_a$, so
$m_a/m_\pi=\sqrt{N_f}f_\pi/2f_a$ is of order $10^{-2}$ across the window and the
ALP is very light compared with the dark pion. Three consequences follow, and
together they leave the branch with very little room.

The first is that $\pi\pi\to aa$ has no threshold. For $m_a\ll m_\pi$ the
conversion is open at every energy the pion bath supplies, and with the contact
amplitude of Eq.~\eqref{eq: portal-cos} it is fast. The requirement that the
$3\to2$ process still dominate the number-changing budget,
\begin{equation}
  \braket{\sigma v^2}_{3\to2}\,n_\pi^2
  > \braket{\sigma v}_{\pi\pi\to aa}\,n_\pi ,
  \label{eq: 3to2-dominance}
\end{equation}
evaluated at the dark freeze-out point $x^\prime=m_\pi/T^\prime\simeq22$, then
demands
\begin{equation}
  f_a \gtrsim 62\,m_\pi ,
  \label{eq: floor-tied}
\end{equation}
which at $m_\pi=140\MeV$ is $f_a\gtrsim8.7\GeV$ and at $600\MeV$ is
$37\GeV$. The second is that the ALP cannot annihilate. Since $m_a\ll m_\pi$ the
reaction $aa\to\pi\pi$ is threshold-suppressed, so the ALPs survive as a hot
relic and decay late, and the resulting spectral distortion and
photodissociation bounds must then be imposed as well.

Neither obstruction fixes a ceiling on $f_a$ by itself, and the third element of
the argument is the one we have not been able to complete. It concerns the relic
ALP.

The ALP is relativistic when the pions freeze out, and it is so whatever the dark
pion mass. Dark freeze-out puts the pions at $T^\prime_f=m_\pi/x^\prime_f$ with
$x^\prime_f\simeq22$, while the tie at the floor puts the ALP at
$m_a=m_\pi/351$, so that
\begin{equation}
  \frac{m_a}{T^\prime_f} \;=\; \frac{x^\prime_f}{351} \;\simeq\; 0.06 ,
  \label{eq: tied-relativistic}
\end{equation}
independently of $m_\pi$. What this does not fix is the ALP number. The elastic
scattering $\pi a\to\pi a$ of Eq.~\eqref{eq: na-over-npi} holds the ALP at the
dark temperature, but it changes no numbers, and the only number-changing channel
available to the ALP is $\pi\pi\leftrightarrow aa$. Chemical equilibrium would
require that channel to supply
$n_a^{\rm eq}/n_\pi\simeq0.38\,e^{x^\prime}/x^{\prime3/2}$ ALPs per pion, which is $1.32\times10^{7}$ at $x^\prime=22$, whereas Eq.~\eqref{eq: 3to2-dominance} sets
$\Gamma_{\pi\pi\to aa}=\Gamma_{3\to2}=H$ at the floor, so that the supply is of
order one per pion per Hubble time. The ALP number therefore decouples long before the pions do. Following the trajectory, supply and demand cross at
    $x^\prime_{\rm cd}=9.2$, the subscript standing for chemical decoupling, where
    $T^\prime=15\MeV$ and $T^\prime/T=0.05$, leaving
\begin{equation}
  Y_a \simeq 7\times10^{-7} ,
  \qquad
  m_a Y_a \simeq 2.9\times10^{-10}\GeV ,
  \label{eq: tied-Ya}
\end{equation}
against $m_\pi Y_\pi=4.4\times10^{-10}\GeV$ for the dark matter itself. The ALP
relic is comparable to the dark matter rather than negligible. At the floor the
tie gives $m_a=m_\pi/351$, which for $m_\pi\lesssim360\MeV$ \textcolor{blue}{} is below $2m_e$
everywhere in the allowed region, so the ALP can only decay to two photons. Its contribution to $N_{\rm eff}$ is negligible, since the frozen number
is $3\times10^{-5}$ of the thermal one.

Whether that relic closes the branch depends on the ALP lifetime $\tau_a$, and hence on the
flavon scale in the tied case, which is the quantity our earlier ceiling depended
on and which the corrected injection integral of Sec.~\ref{sec: relocated} has
not yet been used to supply. What we can say is that the coupled solve does not
reach the observed abundance anywhere the decoupled treatment is valid. At
$m_\pi=140\MeV$, $m_\pi/f_\pi=4$ and $f_a$ at the floor, $\Omega_\pi h^2$ rises as
$V_\phi$ falls but reaches only $0.058$ at the last point where the solve
converges, $V_\phi=2.5\times10^{4}\GeV$, where $T^\prime/T$ has already climbed to
$0.77$. Three times above the floor it reaches $0.096$. Matching the observed
relic therefore requires the two sectors to approach a common temperature, which
is the ordinary freeze-out regime of Fig.~\ref{fig: mesa} and not the decoupled
one this section assumes.

We therefore record the tied realisation as disfavoured rather than excluded, and
name what remains. The floor of Eq.~\eqref{eq: floor-tied} and the kinematics of
Eq.~\eqref{eq: tied-relativistic} are settled, and the relic ALP abundance of
Eq.~\eqref{eq: tied-Ya} is estimated, with a constant $\braket{\sigma v}$ and the
floor normalisation. What is not settled is the flavon scale that the tied
relic condition selects once the corrected injection is used, the resulting
lifetime against the bounds on a decaying relic at this
abundance~\cite{Cadamuro:2011fd,Millea:2015qra}, and the photon coupling in the
short-lifetime corner against the supernova limit on a sub-MeV
ALP~\cite{Carenza:2021ALPe}. Each is a short calculation, and we expect them to
close the branch, but we prefer to leave the statement where the evidence is.

The structural part does not depend on any of this. It follows from the evenness
of Eq.~\eqref{eq: portal-cos}, which removes the trilinear coupling and with it
the cancellation that would have made $\pi\pi\to aa$ slow, and from the tie
itself, which forces $m_a\ll m_\pi$ and so removes both the threshold that would
have made $\pi\pi\to aa$ rare and the one that would have let the ALPs
annihilate away.

\subsection{The generic branch and the flavon scale}

Both obstructions are lifted by the same move. If the explicit term in
Eq.~\eqref{eq: ma-two-pieces} dominates, the ALP mass is free, and once
$m_a>m_\pi$ the conversion $\pi\pi\to aa$ requires $\sqrt s>2m_a$ from a bath of
pions of mass $m_\pi$, so it is threshold- and Boltzmann-suppressed at
freeze-out. The suppression is steep, since $x^\prime\simeq22$,
\begin{equation}
  \braket{\sigma v}_{\pi\pi\to aa} \;\propto\;
  \exp\!\Big[-2x^\prime\Big(\frac{m_a}{m_\pi}-1\Big)\Big] ,
  \label{eq: boltz-supp}
\end{equation}
and the floor falls with it. Table~\ref{tab: floor} collects the result of
imposing Eq.~\eqref{eq: 3to2-dominance} with the contact amplitude.

\begin{table}[t]
  \centering
  \begin{tabular}{|c|c|c|c|c|c|c|}
    \hline
    $m_a/m_\pi$ & $0.01$ & $1.00$ & $1.10$ & $1.20$ & $1.25$ & $1.50$ \\
    \hline
    floor on $f_a$ & $62\,m_\pi$ & $43\,m_\pi$ & $17\,m_\pi$ & $6.2\,m_\pi$
      & $3.7\,m_\pi$ & $0.25\,m_\pi$ \\
    \hline
  \end{tabular}
  \caption{The $3\to2$ dominance floor of Eq.~\eqref{eq: 3to2-dominance} against
  the ALP mass, computed with the contact amplitude that
  Eq.~\eqref{eq: portal-cos} leaves, in the convention of
  Appendix~\ref{app: xsec} in which $|\mathcal{M}|^2$ is quoted per pion-species
  pair. The loss rate is summed over the $N_\pi$ diagonal channels as
  $\sum_b\braket{\sigma v}n_b^2=\braket{\sigma v}n_\pi^2/N_\pi$, with the
  identical-ALP factor $1/2$ in the final state and none in the initial state, and
  is compared with for the total density. The first
  column is the tied case. By $m_a\simeq1.25\,m_\pi$ the floor has
  fallen to a few times the dark pion mass, and above $1.4\,m_\pi$ it drops below
  $m_\pi$ itself and ceases to constrain the model.}
  \label{tab: floor}
\end{table}

We therefore take $m_a\simeq1.25\,m_\pi$, which removes the floor and costs one
parameter. Two further simplifications come with it. Since $2m_a>2m_\pi$ the
reaction $aa\to\pi\pi$ is exothermic and open, so the ALP annihilates into dark
pions rather than surviving, and the hot relic of the tied case does not form.
Moreover $n_a$ is Boltzmann-suppressed relative to $n_\pi$ at freeze-out, so the
dark energy density is carried by the pions alone and the single dark
temperature of Sec.~\ref{sec: cannibal DFO} needs no separate justification.
The spectral distortion, the photodissociation bound on the relic and the
kinetic-decoupling condition all drop out of the analysis together.

With the trilinear coupling absent, the energy that the portal leaks into the
dark sector no longer runs through $f\bar f\to\pi\pi$, which required an
$a\pi\pi$ vertex. It runs instead through the inverse decay $f\bar f\to a$ and
the pair channel $f\bar f\to aa$, both of which use the flavon coupling of
Eq.~\eqref{eq: ALP-SM L} alone. Neither carries the dark scale, and we find that
the injected fraction is independent of $f_a$ to the accuracy of the solve.
The relic condition therefore fixes one number rather than a product,
\begin{equation}
  V_\phi \simeq 1.1\times10^{10}\GeV ,
  \qquad
  \xi \simeq 2.4\times10^{-2}
  \qquad (m_\pi=140\MeV) ,
  \label{eq: relic-vphi}
\end{equation}
and $f_a$ is confined to the window of Eq.~\eqref{eq: fa-window} below. We solve the
coupled Boltzmann system of Sec.~\ref{sec: regimes} directly, with the energy transfer
entering as a source in the $T^\prime$ equation rather than as an initial
temperature ratio, so that no separation between injection and cannibalism has to
be assumed. The value of $V_\phi$ stays within $16\%$ of
$1.15\times10^{10}\GeV$ across a factor of ten in $f_a$, across $m_\pi/f_\pi=2$
to $6$ and across the dark pion mass window, which is the sharpest form of the
statement made in Sec.~\ref{sec: relocated}. In the
kinetically coupled case the relic fixes $f_\pi$ and hence the self-interaction
cross section. Here it fixes the flavon scale outright, and with it every
coupling the ALP has to the Standard Model.

\subsection{The flavon charges and the benchmark}
\label{sec: charges}

The couplings to the visible sector are fixed by the flavon charges, and a
laboratory measurement decides which fermions may carry them. Suppose the flavon
carried the quark hierarchy, with axial charges $c_q$ fixed by the quark Yukawas.
Two consequences follow, and only the second survives at the flavon scale of
Eq.~\eqref{eq: relic-vphi}. The derivative coupling of Eq.~\eqref{eq: ALP-SM L}
first gives the ALP a coupling to nucleons through their axial charges,
\begin{equation}
  g_{aN} = \frac{C_N\, m_N}{V_\phi} ,
  \qquad
  C_N = \sum_{q=u,d,s} c_q\,\Delta q_N ,
  \label{eq: gaN}
\end{equation}
where $\Delta u=0.85$, $\Delta d=-0.41$ and $\Delta s=-0.035$ are the proton spin
fractions~\cite{Raffelt:1996wa}. This coupling has nothing to do with the colour
anomaly and survives even when the anomaly is cancelled exactly. It cannot be
tuned away, because the two combinations
\begin{equation}
  C_p+C_n = (\Delta u+\Delta d)(c_u+c_d) + 2\Delta s\, c_s ,
  \qquad
  C_p-C_n = (\Delta u-\Delta d)(c_u-c_d)
  \label{eq: CpCn}
\end{equation}
cannot both be small, since $\Delta u$ and $\Delta d$ are neither equal nor
opposite, and because the charges are integers with $|c_u|\geq6$ forced by $y_u$.
We have scanned every assignment reproducing the quark and lepton Yukawas with
$\mathcal{O}(1)$ coefficients between $0.1$ and $10$, and the smallest nucleon
coupling any of them reaches is $|C_N|\simeq2.8$, where $|C_N|$ denotes
$\max(|C_p|,|C_n|)$, since the supernova bound constrains whichever of the two is
larger. Narrowing the window on the coefficients to $[0.2,5]$ raises the floor to
$3.7$.

At the flavon scale of Eq.~\eqref{eq: relic-vphi} this gives
$g_{aN}\simeq2.3\times10^{-10}$, which is a factor of $22$ below the SN1987A
energy-loss bound~\cite{Carenza:2019pxu}. The supernova therefore no longer
decides the charge assignment, and at $V_\phi\sim10^{10}\GeV$ the argument has to
be made in the laboratory instead.

The decay $K^+\to\pi^+a$ does make it. An axiflavon carrying the quark hierarchy
leaves a flavour-changing coupling $C_{sd}$ in Eq.~\eqref{eq: ALP-SM L}, since
the PQ charge matrix and the Yukawa matrices are not diagonal in the
same basis, and integrating the derivative coupling by parts turns it into the
scalar density $\bar sd$. The hadronic matrix element is then the $K\to\pi$
scalar form factor, $\braket{\pi^+|\bar sd|K^+}=(m_K^2-m_\pi^2)f_0(q^2)/(m_s-m_d)$,
in which the quark masses cancel against those from the integration by parts, so
that the width is
\begin{equation}
  \Gamma(K^+\to\pi^+a)
  = \frac{|C_{sd}|^2}{4V_\phi^2}\,
    \frac{(m_K^2-m_\pi^2)^2\,|f_0(m_a^2)|^2}{16\pi\,m_K^3}\,
    \lambda^{1/2}(m_K^2,m_\pi^2,m_a^2) ,
  \label{eq: Kpia}
\end{equation}
where $\lambda$ is the K\"all\'en function and the factor $1/4$ is the chiral
projector on the vector current. With $f_0\simeq0.97$ and the generic
Froggatt--Nielsen range $|C_{sd}|\simeq0.1$--$0.22$, Eq.~\eqref{eq: Kpia} gives
$\mathcal{B}(K^+\to\pi^+a)\simeq(0.5$--$2.6)\times10^{-9}$ at the flavon scale of
Eq.~\eqref{eq: relic-vphi}. That range cannot be pushed further down by a
different charge assignment. Reproducing the Cabibbo angle fixes the difference
of the two left-handed doublet charges at one unit, so the left-handed part of
$C_{sd}$ is of order $\epsilon$ whatever the right-handed charges are, and
$|C_{sd}|\simeq0.1$ is a floor rather than a choice.

The ALP is invisible to such a search, since its lifetime of Eq.~\eqref{eq: taua}
exceeds any flight path, so it appears as a peak in the missing mass at
$m_{\rm miss}=m_a$. At $m_a=175\MeV$ that peak lies at
$m_{\rm miss}^2=0.031\GeV^2$, inside the upper signal region
$0.026<m_{\rm miss}^2<0.068\GeV^2$ and well clear of the $\pi^0$ veto that
removes $0.010<m_{\rm miss}^2<0.026\GeV^2$~\cite{NA62:2021zjw}. A dedicated search for $K^+\to\pi^+X$ with a long-lived $X$ covers exactly
    this region. It is performed in two mass intervals, $0<m_X<110\MeV$ and
    $154<m_X<260\MeV$, for lifetimes above $100\,{\rm ps}$, and its strongest
    limit is $\mathcal{B}(K^+\to\pi^+X)<5\times10^{-11}$ at $90\%$ confidence
    for $160<m_X<250\MeV$ and lifetimes above
    $5\,{\rm ns}$~\cite{NA62:2020xlg}. The
benchmark satisfies both conditions, the lifetime of Eq.~\eqref{eq: taua}
exceeding $5\,{\rm ns}$ by eleven orders of magnitude, so the quark-charged
flavon is excluded by a factor of $11$ to $52$. The measured branching ratio
$\mathcal{B}(K^+\to\pi^+\nu\bar\nu)=(9.6^{+1.9}_{-1.8})
\times10^{-11}$~\cite{NA62:2025pnn} says the same thing independently, since it
leaves room for a new contribution only at the level of its own uncertainty,
$\mathcal{B}\lesssim2\times10^{-11}$. The mass window that neither reach covers, $110<m_X<154\MeV$, is the pion
    chimney itself,
which is a further reason to place the benchmark clear of it.
We take the flavon to charge the leptons alone,
\begin{equation}
  c_q = 0 ,
  \qquad
  |c_e|,\,|c_\mu|,\,|c_\tau| = 8,\,5,\,3 ,
  \label{eq: lepto-charges}
\end{equation}
with the charges fixed by the charged-lepton Yukawas through
$y_\ell\sim\epsilon^{|c_\ell|}$ and every $\mathcal{O}(1)$ coefficient between
$0.53$ and $1.18$. The leptophilic assignment gives $C_{sd}=0$ and closes
Eq.~\eqref{eq: Kpia} identically, while the lepton-flavour-violating counterpart
$\mu\to ea$ is kinematically closed at this benchmark, since $m_a$ exceeds
$m_\mu-m_e=105\MeV$.

We note that the exclusion depends on where the ALP mass sits. Searches of this
kind lose their reach in a window of roughly $\pm15\MeV$ around the neutral pion
mass, where the missing mass is degenerate with $K^+\to\pi^+\pi^0$ and where an
ALP mixes resonantly with the $\pi^0$ and decays promptly rather than
escaping~\cite{Balkin:2025koto}. The benchmark below is chosen to sit above that
window, and above the $\pi^0$ veto of the search itself.

\begin{table}[t]
  \centering
  \renewcommand{\arraystretch}{1.6}
  \setlength{\tabcolsep}{18pt}
  \begin{tabular}{|c | c|}
    \hline
     {\bf SM Multiplets} & $U(1)_{\rm PQ}$ {\bf charges} \\
    \hline
    $Q_L^i,\ u_R^i,\ d_R^i$ \ \ (every quark, all generations) & $0$ \\[6pt]
    $\begin{pmatrix}
        \nu_e \\ e
    \end{pmatrix}_L,\ \begin{pmatrix}
        \nu_\mu \\ \mu
    \end{pmatrix}_L,
     \begin{pmatrix}
        \nu_\tau \\ \tau
    \end{pmatrix}_L$ & $0,\ 0,\ 0$\\[6pt]
    $e_R,\ \mu_R,\ \tau_R$ & $8,\ 5,\ 3$\\
    \hline
  \end{tabular}
  \caption{The leptonic Froggatt--Nielsen assignment. $c_f=X_{f_L}+X_{f_R}$ is the
  axial charge that fixes both the ALP--fermion coupling of
  Eq.~\eqref{eq: ALP-SM L} and the Yukawa $y_f\sim\epsilon^{|c_f|}$. No coloured
  field is charged, so $\tfrac12\sum_q c_q=0$ identically and
  $\sum_f c_f N_c^f Q_f^2=16$ comes from the charged leptons alone. With the
  doublets uncharged the ALP--lepton coupling matrix is diagonal on the left, so
  lepton flavour violation is generated only by the right-handed rotation and is
  suppressed by $\epsilon^{|c_i-c_j|}$.}
  \label{tab: charges}
\end{table} The colour anomaly then vanishes not through a cancellation
between charges but because no coloured field carries PQ charge, so
the ALP has no gluon coupling, no nucleon coupling and no hadronic decay channel.
The electromagnetic anomaly is the charged-lepton sum alone,
$E=\sum_f c_f N_c^f Q_f^2=16$. The price is that the quark hierarchy is left to
another mechanism, and we return to that in the conclusions.

We take $m_\pi=140\MeV$ with $f_\pi=35\MeV$, so that $m_\pi/f_\pi=4$ remains
inside the chiral expansion, and $m_a=175\MeV$. The mass is chosen above the
blind window of Sec.~\ref{sec: charges} and below the dimuon threshold
$2m_\mu=211\MeV$, and the self-interaction that follows from it sits inside the
range invoked for small-scale structure. The photon coupling is the fermion
triangle evaluated at the ALP mass rather than the anomaly constant,
$|E_{\rm eff}(m_a)|=9.9$ against $E=16$, since the electron is far lighter than
the ALP and drops out of it. The ALP lifetime is
\begin{equation}
  \tau_a = \frac{8\pi}{g_{ae}^2\,m_a}\Big(1-\frac{4m_e^2}{m_a^2}\Big)^{-1/2} \simeq 
  684 \ {\rm s} ,
  \label{eq: taua}
\end{equation}
where $g_{ae}=c_e m_e/V_\phi$ and the electron pair is the only open channel, the
two-photon width contributing a further three per cent through the triangle.
The dark scale is bounded from above as well as from below, and the upper bound
is not a constraint from data but a condition for the mechanism to work at all.
The energy the portal delivers arrives in ALPs, and it heats the pions only if
the ALPs give it up before the Universe expands. Two channels do that, the
elastic scattering $\pi a\to\pi a$ and the conversion $aa\to\pi\pi$, and both
follow from the same contact coupling $g_{aa\pi\pi}=m_\pi^2/4f_a^2$. Taking the
densities from the solved trajectory rather than from an estimate, and re-solving
the relic condition at each value of the dark scale, the elastic rate per ALP
falls below the expansion rate at $f_a\simeq14\GeV$ and the conversion rate at
$f_a\simeq9\GeV$. The conversion is the one that matters. At the temperature
where the injection peaks the ALPs carry most of their energy as rest mass,
$\braket{E_a}\simeq m_a+3T/2\simeq235\MeV$ against $m_a=175\MeV$, and elastic
scattering can share only the kinetic part. Above $f_a\simeq9\GeV$ the ALPs
therefore keep their rest mass, survive as a separate population with the
lifetime of Eq.~\eqref{eq: taua}, and decay to $e^+e^-$ after nucleosynthesis, so
the ceiling is a physical bound and not merely the edge of validity of the
single-temperature treatment. It is a soft one. The criterion is
$\Gamma/H=1$ with $\Gamma$ falling roughly as $f_a^{-4}$, so a factor of five in
the rate moves the ceiling by only a factor of $1.5$ in $f_a$, and the number
below should be read as an order of magnitude. Together with the floor of
Table~\ref{tab: floor} this leaves
\begin{equation}
  0.5\GeV \;\lesssim\; f_a \;\lesssim\; 9\GeV
  \label{eq: fa-window}
\end{equation}
\begin{figure}[t]
      \centering
      \includegraphics[width=0.9\columnwidth]{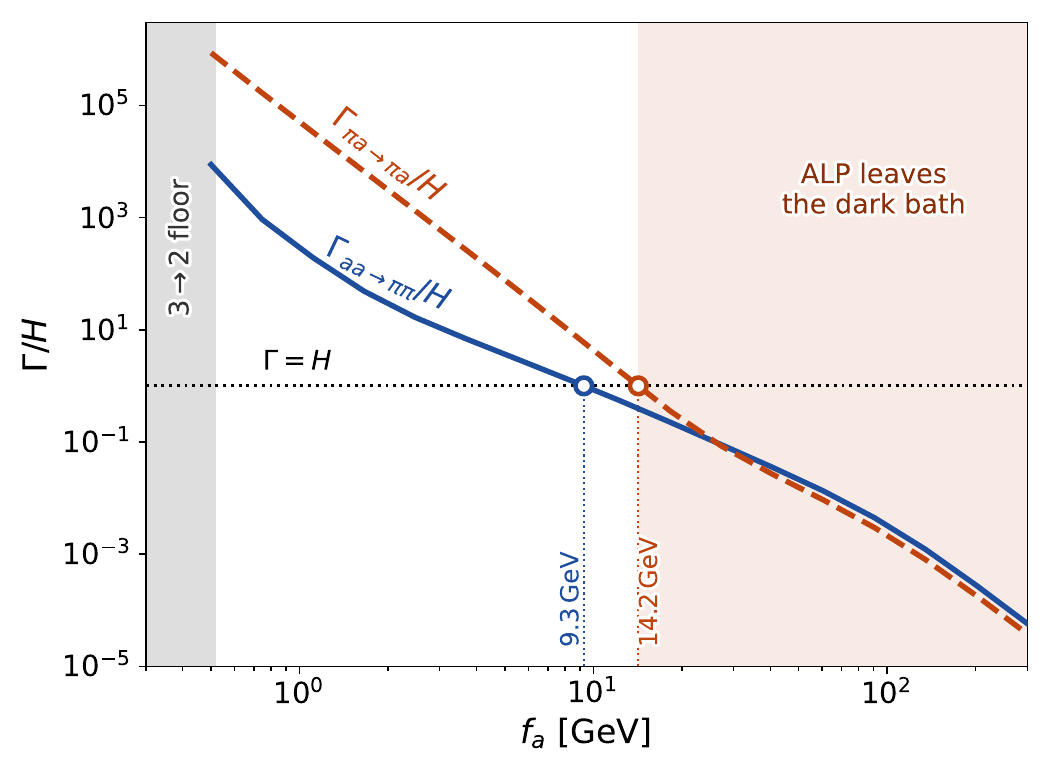}
      \caption{The two channels through which the injected ALPs give their
      energy to the dark pions, against the dark scale, at $m_\pi=140\MeV$,
      $m_a=1.25\,m_\pi$ and $m_\pi/f_\pi=4$. The flavon scale is re-solved for
      $\Omega_\pi h^2=0.12$ at every $f_a$, and each rate is taken from the
      solved trajectory at its maximum rather than from an estimate at a fixed
      temperature. The conversion $aa\to\pi\pi$ falls below the expansion rate
      at $f_a\simeq9.3\GeV$ and the elastic scattering $\pi a\to\pi a$ at
      $14.2\GeV$; the conversion is the one that sets the ceiling, since elastic
      scattering can share only the kinetic part of the ALP energy. The grey
      strip is the $3\to2$ dominance floor of Table~\ref{tab: floor} at this ALP
      mass, $f_a=3.7\,m_\pi$. Together the two give the window of
      Eq.~\eqref{eq: fa-window}.}
      \label{fig: fa-rates}
    \end{figure}
at the benchmark, a window of a little over one decade rather than a half-line. Figure~\ref{fig: fa-rates} shows the two rates across the scan.
The claim that $V_\phi$ does not depend on the dark scale holds inside it, and
Sec.~\ref{sec: relocated} quotes the variation across $f_a=1$ to $10\GeV$ for that reason. Table~\ref{tab: benchmark} collects the resulting predictions.

\textcolor{blue}{
\begin{table}[t]
  \centering
  \begin{tabular}{|l|c|}
  \hline
  \bf{parameters} & \bf{values} \\
    \hline
    $m_\pi$, $f_\pi$ & $140$, $35\MeV$ \\
    $g_{a\gamma}$ & $1.0\times10^{-12}\GeV^{-1}$ \\
    $m_a$ & $175\MeV$ \\ $g_{ae}$ & $3.6\times10^{-13}$ \\
    $V_\phi$ & $1.1\times10^{10}\GeV$ \\ $g_{aN}$ & $0$ \\
    $\xi$ & $2.4\times10^{-2}$ \\ $\tau_a$ & $7.1\times10^{2}\,{\rm s}$ \\
    $f_a$ & $0.5$--$9\GeV$ \\ $\sigma_{\rm self}/m_\pi$
      & $0.20\,{\rm cm^2/g}$ \\
    \hline
  \end{tabular}
  \caption{The benchmark of the generic branch. The ALP decays to $e^+e^-$, the
  channel $a\to\mu^+\mu^-$ being closed at this mass. The flavon scale is a
  prediction of the relic abundance, while the dark scale $f_a$ is bounded from
  below by the $3\to2$ dominance floor of Table~\ref{tab: floor} and from above
  by the condition that the injected ALPs still thermalise with the pions,
  Eq.~\eqref{eq: fa-window}.}
  \label{tab: benchmark}
\end{table}
}

A core-collapse supernova converts thermal photons into ALPs by the Primakoff
process in the screened Coulomb field of the core~\cite{Raffelt:1996wa}. With the
quarks uncharged the nucleon channels are closed, but the electron coupling opens
several others, of which semi-Compton scattering $\gamma e^-\to ae^-$ is the
largest, and the in-medium electron mass reduces the Primakoff rate below its
vacuum estimate~\cite{Fiorillo:2025reloaded}. We compute only the Primakoff
yield below, so the number we quote is a partial count. Both classes of channel
carry $1/V_\phi^2$, so their ratio does not depend on the flavon scale, and at
this benchmark $g_{ae}$ and $g_{a\gamma}\omega$ are comparable at a core energy
$\omega\sim T$. Including the electron channels should therefore raise the yield by an
$\mathcal{O}(1)$ to $\mathcal{O}(10)$ factor. We do not compute it, and we do not
need to: the Primakoff yield below sits more than five orders of magnitude under
the limit, so an enhancement of thirty would still leave four, and the $511\keV$
bound is satisfied whatever the true multiplier turns out to be. The cooling bound on the electron coupling itself is satisfied with
room to spare, since it reaches $g_{ae}\simeq2.5\times10^{-10}$ near
$m_a\simeq120\MeV$~\cite{Carenza:2021ALPe} against the
$3.6\times10^{-13}$ of Table~\ref{tab: benchmark}. For an ALP as heavy as
$175\MeV$ in a core of temperature $T\simeq30\MeV$ the production is drawn from
the tail of the photon distribution and is suppressed by a factor $0.029$
relative to a massless ALP, and the ALPs emerge slower than light, with
$\braket{E_a}\simeq228\MeV$ and $\braket{\gamma\beta}\simeq0.84$.

Two numbers then decide the phenomenology, and they pull in opposite directions.
The decay length is
\begin{equation}
  \lambda = \gamma\beta\,c\,\tau_a \simeq 1.7 \times10^{13}\,{\rm cm} ,
  \label{eq: decay-length}
\end{equation}
which exceeds the radius of the core by seven orders of magnitude and the
    envelope radius of a compact blue-supergiant progenitor,
    $2\times10^{12}\,{\rm cm}$, by a factor of nine. The ALPs therefore leave the star, and the positrons from their decays do
reach the interstellar medium. The Galactic $511\keV$ line is not evaded by
trapping.

It is evaded by production instead. Primakoff conversion carries two powers of the
photon coupling, so the yield falls as $V_\phi^{-2}$, and at the flavon scale of
Eq.~\eqref{eq: relic-vphi} a burst produces 
\begin{equation}
      N_{e^+} = N_{e^+}^{(0)}
        \left(\frac{g_{a\gamma}}{g_{a\gamma}^{(0)}}\right)^{\!2}
        \frac{\mathcal{N}(m_a)}{\mathcal{N}(m_a^{(0)})}
        \simeq 5.6\times10^{47} ,
      \qquad
      \mathcal{N}(m) = \int_m^\infty\! dE\;\frac{E^2}{e^{E/T}-1}
                       \sqrt{1-\frac{m^2}{E^2}} ,
      \label{eq: npos}
    \end{equation}
    where $T\simeq30\MeV$ is the core temperature, $\mathcal{N}$ is the
    Primakoff phase-space weight, and the normalisation
    $(N_{e^+}^{(0)},V_0,m_a^{(0)})=(2.4\times10^{50},\,2.4\times10^{9}\GeV,\,
    7.4\MeV)$ is fixed by reproducing the SN1987A energy-loss bound with the same
    single-zone core. Both powers of $V_\phi$ enter through
    $g_{a\gamma}=(\alpha/2\pi)E_{\rm eff}(m_a)/V_\phi$, whose triangle factor
    $E_{\rm eff}$ also depends on the ALP mass. Even if every positron escaped,
    this lies four orders of magnitude below the limit of $1.4\times10^{52}$
    that the Galactic $511\keV$ measurement
    allows~\cite{Calore:2021lih}. The same suppression
closes the in-flight annihilation continuum.

Since the ALPs leave the star, they deposit nothing in the mantle and the
explosion energetics are untouched~\cite{Lucente:2020whw}. The supernova therefore
places no constraint on this model at all. Its role in the argument is only the
energy-loss bound on the nucleon coupling of Sec.~\ref{sec: charges}, which the
leptonic assignment satisfies identically, since $g_{aN}=0$ there.

The ALP mass controls both the $3\to2$ floor and the decay length, and it moves
them in opposite directions. In the tied case $m_a$ is of order $10^{-2}m_\pi$,
and an ALP that light leaves a $30\MeV$ core ultrarelativistic and long-lived.
Its decay products then reach the Galaxy, and the $511\keV$ line becomes a probe.
That case is disfavoured by the $3\to2$ floor of Sec.~\ref{sec: nogo}, which
follows from the same $m_a\ll m_\pi$. Raising the ALP above the dark pion is what removes
the floor, since it is the Boltzmann suppression of $\pi\pi\to aa$ that lets the
$3\to2$ process dominate. The same step makes the ALP heavy and slow, which
shortens the decay length of Eq.~\eqref{eq: decay-length} sharply, although not
by enough to trap it. What removes the $511\keV$ signal is not the decay length
but the flavon scale, which the relic fixes and which suppresses the Primakoff
yield. The
realisation that survives is therefore the generic branch, with
$m_a\simeq1.25\,m_\pi$ and the couplings of Table~\ref{tab: benchmark}. It is
viable, and what it gives up is the MeV sky rather than the model. Its prediction
lies in the dark sector instead, and we take it up in Sec.~\ref{sec: pheno}.

%% file: sec_pheno.tex

\section{Constraints and phenomenology}
\label{sec: pheno}

Section~\ref{sec: tied} left the dark scale $f_a$ free and fixed the flavon scale
at $V_\phi\simeq1.1\times10^{10}\GeV$. We now collect the constraints on that
prediction. We find that the laboratory and astrophysical bounds on the ALP are
all satisfied by wide margins. The cosmological constraints are absent rather than
merely weak, and what remains as a probe is the dark matter self-interaction.

In the decoupled branch the relic no longer determines
$\sigma_{\rm self}/m_\pi$, which is the content of Sec.~\ref{sec: relocated}. It
is instead determined by the dark pion mass and the ratio $m_\pi/f_\pi$,
\begin{equation}
  \frac{\sigma_{\rm self}}{m_\pi}
  = \frac{1}{32\pi\,m_\pi^3}\left(\frac{m_\pi}{f_\pi}\right)^{\!4} ,
  \label{eq: sigma-self}
\end{equation}
where the third power of the mass is what matters. Equation~\eqref{eq: sigma-self}
is quoted in the normalisation of Ref.~\cite{Hochberg:2014kqa} for a single
lightest pion, in which the coset factor is $a^{\prime2}=1$ for $Sp$ and $4$ for
$SU$ or $SO$. For the degenerate multiplet of Sec.~\ref{sec: model} the same
reference carries instead a factor $a^2/N_\pi^2$, which we have not evaluated, so
the numbers below should be read with an $\mathcal{O}(1)$ coset uncertainty that
does not affect the shape of the window. This is the same relation that the
canonical SIMP mechanism carries, but it is no longer tied to the relic, so the
two ends of the window are set by different physics. At the benchmark it
gives
\begin{equation}
  \frac{\sigma_{\rm self}}{m_\pi} \simeq 0.20\,{\rm cm^2/g} ,
  \label{eq: sigma-bench}
\end{equation}
which sits inside the range $0.1$--$1\,{\rm cm^2/g}$ in which self-interaction is
invoked to address the small-scale structure of haloes~\cite{Tulin:2017ara}. The
bullet cluster~\cite{Randall:2007ph} bounds the same quantity from above and so
bounds the dark pion mass from below, and at $m_\pi/f_\pi=4$ we find
\begin{equation}
  m_\pi \gtrsim 82\MeV ,
  \label{eq: bullet-floor}
\end{equation}

\begin{figure}[t]
  \centering
  \includegraphics[width=0.86\textwidth]{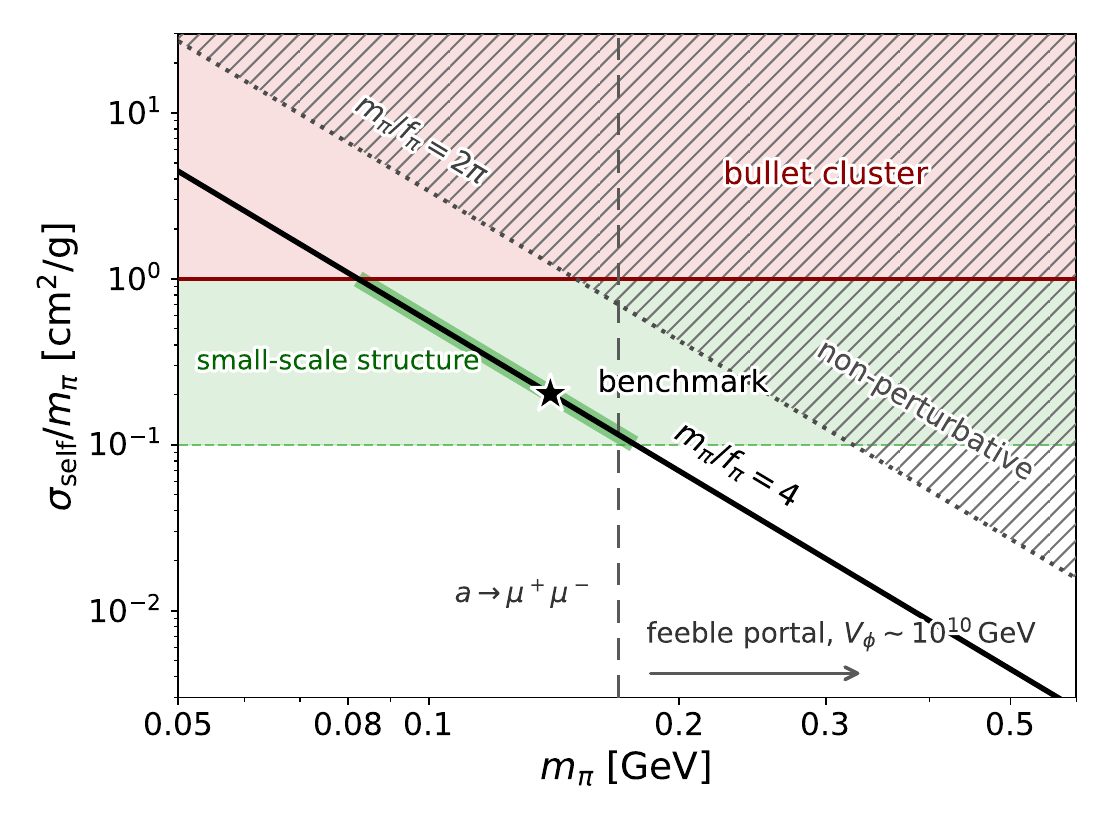}
  \caption{The self-interaction plane. The green band is the range
  $0.1$--$1\,{\rm cm^2/g}$ in which self-interaction is invoked to address
  small-scale structure, and the red region above it is excluded by the bullet
  cluster~\cite{Randall:2007ph}, which bounds $\sigma_{\rm self}/m_\pi$ rather
  than the dark pion mass. The hatched region lies above the chiral
  perturbativity edge $m_\pi/f_\pi=2\pi$. The solid line is
  Eq.~\eqref{eq: sigma-self} at the ratio $m_\pi/f_\pi=4$ adopted throughout,
  thickened where it crosses the favoured band, which it does between
  $82\MeV$ and $177\MeV$. Along the perturbativity edge itself the bullet
  cluster is reached instead at $m_\pi\simeq150\MeV$. The vertical dashed line is
  the kinematic threshold at which $a\to\mu^+\mu^-$ opens, and it is a reference
  rather than an exclusion. Above it the inverse decay through muons dominates the
  energy injection, and since the muon carries $\Gamma_{\mu\mu}/\Gamma_{ee}
  \simeq2\times10^{4}$ the flavon scale that the relic selects jumps by nearly two
  orders of magnitude, from $1.1\times10^{10}$ to $5\times10^{11}\GeV$ at the
  threshold and to $10^{12}\GeV$ shortly above it. That leaves the model viable
  but even further below every laboratory reach. The star marks the benchmark of Sec.~\ref{sec: tied},
  $m_\pi=140\MeV$ and $\sigma_{\rm self}/m_\pi=0.20\,{\rm cm^2/g}$.}
  \label{fig: sidm window}
\end{figure}
while the upper end of the favoured band is reached at $m_\pi\simeq177\MeV$.
Figure~\ref{fig: sidm window} collects this. Along the perturbativity edge
$m_\pi/f_\pi=2\pi$ the bullet cluster is reached at $m_\pi\simeq150\MeV$ instead,
so the bound on the dark pion mass depends on the chiral ratio and is quoted here
at the value we adopt. The
model therefore predicts a self-interaction that is observable, and it does so
across essentially the same mass range that Sec.~\ref{sec: tied} selects on
entirely independent grounds. Above $m_\pi\simeq169\MeV$ the ALP mass crosses
$2m_\mu$, the muonic decay opens, and the flavon scale that the relic selects
jumps by two orders. The window in $m_\pi$ therefore closes from above at almost
the same place that Eq.~\eqref{eq: sigma-self} leaves the favoured band. That the two
coincide is a coincidence of the numbers rather than a structural result, but it
is what makes the surviving parameter space narrow.

The cosmological bounds that closed the tied case are absent here rather
than weak.
Since $m_a>m_\pi$ the reaction $aa\to\pi\pi$ is exothermic, so the ALP
annihilates into dark pions and no hot relic survives to decay. There is
therefore no injection into the cosmic microwave background and no chemical
potential distortion, and the photodissociation bound of
Ref.~\cite{Forestell:2018txr} has nothing to act on. The dark sector is also
much colder and much thinner than the visible one. At $T=1\MeV$ the solved
trajectory gives $\rho^\prime/\rho_{\rm SM}=6.8\times10^{-7}$, with the dark
pions already deeply non-relativistic, $m_\pi/T^\prime\simeq4\times10^2$, so they
do not contribute to $N_{\rm eff}$ at all and the bound is on the energy density,
which is met by six orders of magnitude. The dark pions become non-relativistic during
the cannibal phase and are heavy enough that the Lyman-$\alpha$
bound~\cite{Irsic:2017ixq} is comfortably met.

Because the flavon charges the leptons and nothing else, the laboratory
constraints that matter are leptonic. The predicted electron coupling is
$g_{ae}\simeq3.6\times10^{-13}$, which lies five orders of magnitude below the
weak-coupling edge of the E137 exclusion, where the sensitivity closes because
the ALP is produced too rarely rather than because it decays too
soon~\cite{Dolan:2017osp}. Missing-energy searches at NA64 reach an effective
coupling four orders above the prediction. The laboratory programme and the
model do not overlap.

The lepton-flavour-violating decay $\mu\to e\,a$, which is the leptonic
counterpart of the meson decays that constrain a quark flavon, is
\emph{kinematically closed} at this benchmark, since $m_a=175\MeV$ exceeds
$m_\mu-m_e$. It reopens only for $m_a\lesssim105\MeV$, that is
$m_\pi\lesssim84\MeV$, which Eq.~\eqref{eq: bullet-floor} already restricts to a
narrow sliver at the bottom of the allowed mass range. Should the dark pion mass
lie there, the limit on a light ALP in muon decay,
$V_\phi/|c_{\mu e}|\gtrsim5.5\times10^{9}\GeV$~\cite{Calibbi:2020jvd}, with
$c_{\mu e}\sim|c_e-c_\mu|\epsilon^{|c_e-c_\mu|}\simeq3\times10^{-2}$, would
require $V_\phi\gtrsim1.8\times10^{8}\GeV$, which the prediction of
Eq.~\eqref{eq: relic-vphi} clears by nearly two orders. The tau channels are
weaker by four orders again. At the flavon scale the relic now selects, no
laboratory search reaches this model, and the probe that remains is the dark
sector itself.

The kaon argument of Sec.~\ref{sec: charges} is what forces the leptonic
assignment, and the leptonic model itself sits far below every supernova bound on
the photon coupling. The ALPs that a supernova produces do leave the star at this
flavon scale, but Primakoff conversion carries two powers of the photon coupling,
so the yield falls as $V_\phi^{-2}$ and a burst releases only
$7.1\times10^{46}$ positrons. The Galactic $511\keV$ line and the in-flight
annihilation continuum therefore place no constraint on this model. We regard
that as a loss rather than a success, since it is the observational channel that
the tied realisation would have offered had it been viable, and
Sec.~\ref{sec: tied} explains why no version of the model can have both.

What the relic abundance predicts here is a flavon scale, and what tests it is
the dark sector rather than the portal. The self-interaction of
Eq.~\eqref{eq: sigma-bench} lies in the range that halo shapes, cluster mergers
and the diversity of rotation curves already probe. An improvement in the
inferred $\sigma_{\rm self}/m_\pi$ of a factor of a few, in either direction,
could therefore select or exclude the dark pion mass range that this model
requires.
A measurement below $0.1\,{\rm cm^2/g}$ would push $m_\pi$ above $177\MeV$, where
the muonic decay opens and the flavon scale jumps, and a measurement at
$1\,{\rm cm^2/g}$ would put the model at the bullet cluster edge. That is a
narrower target than the canonical SIMP window, and it is reached from the
opposite direction, since here the self-interaction is an output of the surviving
parameter space rather than a consequence of the relic condition.

%% file: sec_conclusions.tex

\section{Conclusions}
\label{sec: conclusions}

In this work, we have revisited SIMP dark matter through the axion portal and
mapped its cosmology as a function of the ALP mass and the portal coupling. The
picture that emerges is a phase diagram with a freeze-in region, a decoupled
region and a freeze-out region, and we find that the observed relic sits in the
decoupled region for any viable portal.

We began by asking whether the $3\to2$ process really requires a heavy ALP, and
we find that it does not. The process dominates the hidden annihilation
$\pi\pi\to aa$ whenever the ALP decay constant exceeds the pion decay constant,
for any ALP mass, so the cosmology is selected by the portal strength and not by
the kinematics. We then analysed kinetic contact with the SM and found that it is
carried by the elastic $\pi a\to\pi a$ scattering and by the off-shell
$\pi f\to\pi f$ scattering, both of which survive only for a strong portal in the
range already probed by beam-dump and rare-decay searches. Once the portal is
weak enough to be viable, the dark sector decouples for any ALP mass.

We then solved the cannibal freeze-out in the decoupled sector and find that the
relic follows $\Omega_\pi h^2\propto(s^\prime/s)\,T^\prime_f$, with the $3\to2$
coupling entering only logarithmically. The energy the portal delivers arrives
after the dark bath has become non-relativistic, so the entropy it deposits is
linear in it and the relic is quadratic in the portal coupling up to that
logarithm. The relic therefore no longer predicts the self-interaction cross
section, and the canonical SIMP mass window opens. Since the injected entropy is
generated by the energy that the portal leaks from the SM, the relic instead
predicts the ALP--SM coupling. This is what we mean by relocating the miracle.

We then imposed the minimal realisation, in which the ALP is the flavon and its
mass comes from the portal itself, and here we find a structural obstruction that
an effective trilinear treatment obscures. The portal operator carries its Hermitian
conjugate, and the invariant it contains is real, so the operator is even in the
ALP field. No trilinear $a\pi\pi$ coupling is generated, the dark sector couples
to the ALP only in pairs, and the conversion $\pi\pi\to aa$ proceeds through a
single contact diagram. The near-cancellation that a trilinear coupling would
have supplied is absent, and we find the conversion to be some four orders of
magnitude faster than a trilinear estimate would give.

That is what disfavours the tied-mass realisation. With the condensate alone
generating the mass, the ALP is two orders lighter than the dark pion, so the
conversion carries no threshold and the requirement that $3\to2$ dominate demands
$f_a\gtrsim62\,m_\pi$. The same tie fixes $m_a/T^\prime_f\simeq0.06$ whatever the
dark pion mass, so the ALP is relativistic when the pions freeze out and survives
as a relic comparable to the dark matter itself, which can only decay to photons.
We find in addition that the coupled solve does not reach the observed abundance
anywhere the two sectors remain decoupled. What we have not done is fix the
flavon scale in this branch and confront the resulting lifetime with the bounds
on a decaying relic, and we record the branch as disfavoured rather than
excluded until that is done. We take the generic branch in what follows.

The realisation that survives keeps the flavon but abandons the tie. Once the ALP
is slightly heavier than the dark pion, the conversion requires
$\sqrt s>2m_a$ from a bath of lighter pions, so it is Boltzmann suppressed at
freeze-out and we find that $m_a\simeq1.25\,m_\pi$ removes the floor entirely.
Moreover $aa\to\pi\pi$ is then exothermic, so the ALP annihilates into dark pions
and no hot relic forms, which removes the spectral distortion and the
photodissociation bound together. The energy that the portal leaks no longer runs
through a trilinear vertex, so it does not carry the dark scale, and matching
$\Omega_\pi h^2=0.12$ fixes the flavon scale outright at
$V_\phi\simeq1.1\times10^{10}\GeV$ while leaving $f_a$ free. This is the sharpest
form of the relocated miracle, since the relic now predicts a single number
rather than a product of two.

We find that the supernova decides which fermions the flavon may charge, and that
it does so through energy loss rather than through what the ALPs do afterwards. A
flavon carrying the quark hierarchy gives the ALP a coupling to nucleons through
their axial charges, and a flavour-changing coupling that opens $K^+\to\pi^+a$.
At the flavon scale that the relic selects the nucleon coupling falls below the
SN1987A bound, but the kaon decay exceeds the NA62 limit on $K^+\to\pi^+X$
by one to two orders of magnitude. The surviving realisation charges the
leptons alone, and its colour anomaly then vanishes identically rather than by
cancellation.

Finally, we followed the ALPs that a core-collapse supernova produces, and here
the model loses an observable and gains another. At $175\MeV$ the ALP emerges
only mildly relativistic from a $30\MeV$ core, and the flavon scale that the
relic selects makes it long-lived. Its decay length of
$1.8\times10^{13}\,{\rm cm}$ leaves the core easily and exceeds the smallest
progenitor envelope by a factor of nine. The ALPs therefore leave the star, and what
protects the Galactic $511\keV$ line is the Primakoff yield instead, which falls
as $V_\phi^{-2}$ and lands five orders below the limit. We want to remark that this is not a feature of the
benchmark. The ALP mass sets both the $3\to2$ floor and the decay length, and it
moves them in opposite directions. What the surviving branch gives up is therefore
the MeV sky rather than the model itself.

What replaces it is the self-interaction. Since the relic no longer fixes
$\sigma_{\rm self}/m_\pi$, it is set by the dark pion mass alone, and at
$m_\pi=140\MeV$ we find $0.20\,{\rm cm^2/g}$, inside the range invoked to address
small-scale structure. The bullet cluster bounds the dark pion mass from below at
$82\MeV$, while above $169\MeV$ the ALP crosses the dimuon threshold and the
flavon scale jumps by two orders. The surviving window is therefore narrow, and
it is probed by halo shapes and cluster mergers rather than by MeV astronomy.

Several limitations should be stated. The lepton charges and the hierarchy
$f_a\ll V_\phi$ have been adopted as inputs, and the leptonic assignment leaves
the quark hierarchy to a separate mechanism, which we have not constructed. The
supernova luminosity is computed in a single-zone core and normalised so that the
same machinery reproduces the SN1987A energy-loss bound, which fixes its overall
scale only to within a factor of two. The conclusion that the ALPs leave the star
survives that uncertainty, since the decay length exceeds the envelope by a
factor of nine. The coincidence between the self-interacting window and the dimuon
threshold is a coincidence of the numbers and not a structural result. Moreover
the choice $N_c=2$ against $N_c=4$ remains an unresolved convention, and it
shifts $\braket{\sigma v^2}_{3\to2}$ and the floor of Table~\ref{tab: floor} by a
factor of four. Finally, the dark scale $f_a$ is confined to about one decade rather than
predicted, bounded below by the $3\to2$ dominance floor and above by the
requirement that the injected ALPs thermalise with the pions, and a second
observable would be needed to fix it within that window.

Several directions remain. The production side should be completed with the
radiative and gauge-boson channels, which could shift the predicted flavon scale.
The self-interaction is fixed by the dark pion mass, with no free normalisation
left to adjust. An improvement of a factor of a few in the inferred
$\sigma_{\rm self}/m_\pi$, in either direction, could therefore select or exclude
the mass range this model requires. It would also be interesting to sharpen the
predicted flavon scale with precision cosmological observables, since the relic
condition is what fixes it. We leave such studies for possible future work.

%% file: appendix.tex
%

\section{Chiral Lagrangian, WZW term and the portal expansion}
\label{app: portal}

This appendix collects the effective Lagrangian of the dark sector and the full
expansion of the portal operator quoted in Eq.~\eqref{eq:ALP-DM L}.

The $Sp(2N_f)$ invariant and the pion field are
\begin{equation}
  J = \begin{pmatrix} 0 & \mathbf{1}_{N_f} \\ -\mathbf{1}_{N_f} & 0 \end{pmatrix} ,
  \qquad
  \Sigma = \exp\!\big(2i\pi/f_\pi\big)\,J ,
\end{equation}
where $J^2=-\mathbf{1}_{2N_f}$, so that
${\rm Tr}[J\Sigma]={\rm Tr}[J\,e^{2i\pi/f_\pi}J]={\rm Tr}[J^2 e^{2i\pi/f_\pi}]
=-{\rm Tr}[e^{2i\pi/f_\pi}]$.
Because $\pi$ is Hermitian and traceless and the coset satisfies
${\rm Tr}[\pi^3]=0$, the Cayley--Hamilton theorem for a traceless $4\times4$ matrix
gives ${\rm Tr}[\pi^{2k+1}]=0$ for every odd power, so ${\rm Tr}[e^{2i\pi/f_\pi}]$
is real. The Hermitian conjugate in Eq.~\eqref{eq: portal-op} therefore combines
the two exponentials into a cosine,
\begin{align}
  \mathcal{L}_{a\pi}
  &= -\tfrac12 m_{\rm PQ}^2 a^2
     + \tfrac12 m_Q\mu^3\Big(e^{ia/f_a}\,{\rm Tr}\big[e^{2i\pi/f_\pi}\big]
       + {\rm h.c.}\Big) \nonumber\\
  &= -\tfrac12 m_{\rm PQ}^2 a^2
     + m_Q\mu^3\,\cos\!\Big(\frac{a}{f_a}\Big)\,
       {\rm Tr}\Big[\cos\!\Big(\frac{2\pi}{f_\pi}\Big)\Big] ,
  \label{eq: app-expansion}
\end{align}
which is even in $a$ and even in $\pi$ separately. This is the central structural
point of the portal. Retaining only one exponential would generate the odd vertices
$a\pi\pi$, $a^3$ and $a^3\pi\pi$, each carrying an explicit factor of $i$; every one
of them cancels against its conjugate. Expanding both cosines and keeping the terms
with at most four legs gives
\begin{align}
  \mathcal{L}_{a\pi} \supset\;
  & \left(-\tfrac12 m_{\rm PQ}^2 - \frac{N_f m_Q\mu^3}{f_a^2}\right)a^2
    - \frac{2 m_Q\mu^3}{f_\pi^2}{\rm Tr}[\pi^2]
    + \frac{m_Q\mu^3}{f_\pi^2 f_a^2}\,a^2\,{\rm Tr}[\pi^2] + \cdots ,
\end{align}
where the ellipsis stands for terms of higher order in $\pi/f_\pi$ and $a/f_a$, all
of them even in each. The odd invariants ${\rm Tr}[\pi]$ and ${\rm Tr}[\pi^3]$ drop
out independently, so no cubic pion vertex survives either.

For the $SU(4)/Sp(4)$ coset the pion matrix is
\begin{equation}
  \pi =
  \begin{pmatrix}
    \frac{\pi_3}{\sqrt2} & \frac{\pi_1 - i\pi_2}{\sqrt2} & 0
      & \frac{\pi_4 - i\pi_5}{\sqrt2} \\
    \frac{\pi_1 + i\pi_2}{\sqrt2} & -\frac{\pi_3}{\sqrt2}
      & -\frac{\pi_4 - i\pi_5}{\sqrt2} & 0 \\
    0 & -\frac{\pi_4 + i\pi_5}{\sqrt2} & \frac{\pi_3}{\sqrt2}
      & \frac{\pi_1 + i\pi_2}{\sqrt2} \\
    \frac{\pi_4 + i\pi_5}{\sqrt2} & 0 & \frac{\pi_1 - i\pi_2}{\sqrt2}
      & -\frac{\pi_3}{\sqrt2}
  \end{pmatrix} ,
  \label{eq: pion matrix}
\end{equation}
from which the two identities used above follow, namely ${\rm Tr}[\pi^3]=0$ and
${\rm Tr}[\pi^2]=2\pi^a\pi^a$. Identifying the dark pion mass through
$m_\pi^2/8=m_Q\mu^3/f_\pi^2$, the Hermitian conjugate having doubled every surviving
coefficient, and setting $N_f=2$ then gives the compact form quoted as
Eq.~\eqref{eq:ALP-DM L} in the main text. The tied mass
$m_a^2=N_f m_\pi^2 f_\pi^2/4f_a^2$ and the contact coupling $m_\pi^2/4f_a^2$ are
unchanged by the doubling, since it is absorbed into the identification of
$m_\pi$.

The Wess--Zumino--Witten term of Eq.~\eqref{eq:WZW term} reduces for five pions to
a single five-point vertex,
\begin{equation}
  \mathcal{L}_{\rm WZW}
  = \frac{8}{15\pi^2 f_\pi^5}\,\epsilon^{\mu\nu\rho\sigma}\,T_{12345}\,
    \pi_1\partial_\mu\pi_2\partial_\nu\pi_3\partial_\rho\pi_4\partial_\sigma\pi_5 ,
\end{equation}
whose Feynman rule, together with those from Eqs.~\eqref{eq:ALP-DM L},
\eqref{eq: ALP-SM L} and \eqref{eq: gag-total}, is collected in
Table~\ref{tab: feynman rules}.

\begin{table}[h]
  \centering
  \renewcommand{\arraystretch}{2.0}
  \begin{tabular}{|ccccc|l|}
    \hline
    \multicolumn{5}{|c|}{Particles} & \multicolumn{1}{c|}{Vertex factor} \\
    \hline
    $\pi^b$ & $\pi^c$ & $a$ & $a$ & & $\displaystyle i\,\frac{m_\pi^2}{f_a^2}\,\delta^{bc}$ \\
    $f$ & $\bar f$ & $a$ & & & $\displaystyle -\,\frac{c_f m_f \gamma_5}{V_\phi}$ \\
    $\gamma$ & $\gamma$ & $a$ & & & $ig_{a\gamma}\,\epsilon^{\mu\nu\rho\sigma}p_{1\sigma}p_{2\rho}$ \\
    $\pi_1$ & $\pi_2$ & $\pi_3$ & $\pi_4$ & $\pi_5$ &
      $\displaystyle \frac{8}{15\pi^2 f_\pi^5}\,T_{12345}\,
       \epsilon^{\mu\nu\rho\sigma}p_{2\mu}p_{3\nu}p_{4\rho}p_{5\sigma}$ \\
    \hline
  \end{tabular}
  \caption{Vertex factors following from Eqs.~\eqref{eq:ALP-DM L},
  \eqref{eq: ALP-SM L}, \eqref{eq: gag-total} and \eqref{eq:WZW term}, with the
  $af\bar f$ rule following from Eq.~\eqref{eq: ALP-SM L} as
  $i\times i c_f m_f\gamma_5/V_\phi$, so that it is real and negative, and reducing
  to the single-scale case on setting $V_\phi=f_a$. The
  $aa\pi\pi$ entry is the vertex factor rather than the Lagrangian
  coefficient of Eq.~\eqref{eq:ALP-DM L}, and so carries the multiplicity $4$
  of the functional derivative, in agreement with
  Eq.~\eqref{eq: vertex-rules}. There is no
  $a\pi\pi$ and no $a^3$ vertex: the portal of Eq.~\eqref{eq: app-expansion} is
  even in $a$, so the dark sector couples to the ALP only in pairs. The $a\gamma\gamma$
  rule follows from $-\tfrac14 g_{a\gamma}aF_{\mu\nu}\tilde F^{\mu\nu}$ with
  $\tilde F_{\mu\nu}=\tfrac12\epsilon_{\mu\nu\rho\sigma}F^{\rho\sigma}$, and for the
  five-pion vertex all momenta flow into the vertex.}
  \label{tab: feynman rules}
\end{table}

\section{Cross sections and squared amplitudes}
\label{app: xsec}

This appendix lists the squared amplitudes for every $2\to2$ process entering the
Boltzmann system of Sec.~\ref{sec: regimes}. They have been obtained with
\texttt{FeynCalc}. Each is summed over the spins and colours of both the initial
and the final state, so that the initial-state average, $1/4N_c^2$ for a fermion
pair, is restored when the cross section is formed, and the ALP propagator carries
a Breit--Wigner width $\Gamma_a$ wherever the ALP can go on shell in the $s$
channel. Throughout, $u$ is fixed by $s+t+u=\sum_i m_i^2$, and the visible
couplings are written on the single scale $f_a$ of the generic case, so that
$f_a\to V_\phi$ in every $af\bar f$ and $a\gamma\gamma$ vertex restores the
two-scale form of Sec.~\ref{sec: model}. The differential cross section follows
from the squared amplitude through
\begin{equation}
  \left(\frac{d\sigma}{d\Omega}\right)_{\rm CM}
  = \frac{1}{64\pi^2 s}\,
    \frac{\sqrt{\lambda(s,m_3^2,m_4^2)}}{\sqrt{\lambda(s,m_1^2,m_2^2)}}\,
    |\mathcal{M}|^2 ,
  \label{eq: dsigma}
\end{equation}
where $\lambda$ is the usual K\"all\'en function.

We keep the dark pion decay constant $f_\pi$ explicit, since it is an independent
parameter in the generic case of
Sections~\ref{sec: regimes}--\ref{sec: relocated}. It enters nowhere except
through the $a^3$ vertex of Eq.~\eqref{eq:ALP-DM L}, and therefore only through
$s$-channel ALP exchange, which for $f_a\gg f_\pi$ is suppressed relative to the
$t$- and $u$-channel pieces by $f_\pi^2/f_a^2$. In the minimal case the tie of
Eq.~\eqref{eq: tied-min} removes it through
$f_\pi=2m_af_a/\sqrt{N_f}\,m_\pi$, and every expression below collapses to a
function of $m_\pi$, $m_a$ and $f_a$ alone.

\subsection*{Portal processes with two pions}

The reaction $\pi\pi\to\gamma\gamma$ proceeds through a single $s$-channel ALP, and
with $k_{1,2}$ the photon momenta the squared amplitude is
\begin{equation}
  |\mathcal{M}|^2_{\pi\pi\to\gamma\gamma}
  = \tan^2\!\bar\theta\;
    \frac{g_{a\gamma}^2\, m_\pi^4\, s^2}
         {8 f_a^2\big(m_a^4 + m_a^2(\Gamma_a^2-2s) + s^2\big)} .
  \label{eq: pipi-gaga}
\end{equation}
The reverse process $\gamma\gamma\to\pi\pi$ has the same squared amplitude, as
required by detailed balance at this order.

The reaction $\pi\pi\to f\bar f$ likewise proceeds through a single $s$-channel ALP,
and with $p_{1,2}$ the fermion momenta the squared amplitude is
\begin{equation}
  |\mathcal{M}|^2_{\pi\pi\to f\bar f}
  = \tan^2\!\bar\theta\;
    \frac{c_f^2\, m_f^2\, m_\pi^4\, s}
         {2 f_a^4\big(m_a^4 + m_a^2(\Gamma_a^2-2s) + s^2\big)} ,
  \label{eq: pipi-ff}
\end{equation}
with $|\mathcal{M}|^2_{f\bar f\to\pi\pi}$ equal to it. Both amplitudes run through
a single $a\pi\pi$ vertex, so both carry the factor $\tan^2\!\bar\theta$ defined
below in Eq.~\eqref{eq: theta-vertices} and both vanish identically at
$\bar\theta=0$, which is where this model sits. In a portal that did
carry the trilinear this would be the channel that dominates the energy injection
of Sec.~\ref{sec: relocated}. With it closed, the injection runs through the
inverse decay and through Eq.~\eqref{eq: ffaa-tu} instead. The elastic scattering
$\pi f\to\pi f$ used in Sec.~\ref{sec: consistency} is obtained by crossing
Eq.~\eqref{eq: pipi-ff} into the $t$ channel, which gives Eq.~\eqref{eq: pif} of
the main text.

\subsection*{Portal processes with two ALPs}

The reaction $f\bar f\to aa$ receives one $s$-channel contribution through the
$a^3$ vertex and two contributions with an internal fermion in the $t$ and $u$
channels. The $s$-channel piece,
\begin{equation}
  |\mathcal{M}|^2_{f\bar f\to aa}\big|_{s}
  = \frac{c_f^2 f_\pi^4 m_f^2 m_\pi^4 s}{72 f_a^8 (m_a^2-s)^2} ,
  \label{eq: ffaa-s}
\end{equation}
is suppressed by $f_\pi^4/f_a^8$ and is negligible for $f_a\gg f_\pi$. The fermion
exchange uses the $af\bar f$ vertex twice, so it carries four powers of the portal
coupling and no $f_\pi$ at all,
\begin{align}
  |\mathcal{M}|^2_{f\bar f\to aa}\big|_{tu}
  &= \frac{2\, c_f^4\, m_f^4}{f_a^4\,(m_f^2-t)^2(m_f^2-u)^2}\,
     \Big[ m_a^4\big(t+u-2m_f^2\big)^2 \nonumber\\
  &\qquad + m_a^2\big(8m_f^6 - s(t-u)^2 - 12m_f^4(t+u)
       + 6m_f^2(t+u)^2 - (t+u)^3\big) \nonumber\\
  &\qquad + (m_f^2-t)(m_f^2-u)\big(4m_f^4 - s^2 - 4m_f^2(t+u)
       + 2(t^2+u^2)\big)\Big] ,
  \label{eq: ffaa-tu}
\end{align}
where $u=2m_f^2+2m_a^2-s-t$. It is this piece that drives the $2\to2$ energy
injection of Sec.~\ref{sec: tied}, and it is the origin of the four powers of
$g_{aff}$ in Eq.~\eqref{eq: xi-scaling-closed}. The reverse reaction
$aa\to f\bar f$ has the same squared amplitude, as time-reversal invariance
requires, so the two differ only by the initial-state average.

The reaction $\gamma\gamma\to aa$ proceeds through the $s$-channel $a^3$ vertex
and through $t$- and $u$-channel photon exchange, with the result
\begin{align}
  |\mathcal{M}|^2_{\gamma\gamma\to aa}
  &= \frac{f_\pi^4 g_{a\gamma}^2 m_\pi^4 s^2}{288 f_a^6 (m_a^2-s)^2} \nonumber\\
  &\quad + \frac{g_{a\gamma}^4}{8t^2u^2}
     \Big[4m_a^{12} - 4m_a^{10}(s+4t) + m_a^8(s^2+24st+24t^2) \nonumber\\
  &\qquad - 4m_a^6 t(3s^2+10st+4t^2)
     + 2m_a^4 t(s^3+21s^2t+12st^2+2t^3) \nonumber\\
  &\qquad - 4m_a^2 st^2(6s^2+7st+t^2) + 5s^2t^2(s+t)^2\Big] ,
  \label{eq: gg-aa}
\end{align}
where $u=2m_a^2-s-t$ for massless photons, and
$|\mathcal{M}|^2_{aa\to\gamma\gamma}$ is equal to it.

\subsection*{The hidden-sector conversion and the elastic channel}

These two amplitudes follow from the portal of Eq.~\eqref{eq: app-expansion}, and
their structure is controlled by whether the ALP vacuum conserves CP. In this
model it does, for the reason given below Eq.~\eqref{eq: portal-op}, and the odd
vertices are absent throughout. We nonetheless record what they would be in a
portal that possessed them, both as a check that the amplitudes below reduce
correctly and because the trilinear is what a treatment through an effective
$a\pi\pi$ coupling would retain. Writing $\bar\theta$ for the effective angle at
the minimum of the ALP potential, which is zero here, the Lagrangian couplings
are
\begin{equation}
  \frac{m_\pi^2}{4f_a^2}\,a^2\pi^a\pi^a ,
  \qquad
  \tan\bar\theta\,\frac{m_\pi^2}{2f_a}\,a\,\pi^a\pi^a ,
  \qquad
  \tan\bar\theta\,\frac{N_f m_\pi^2 f_\pi^2}{24 f_a^3}\,a^3 ,
  \label{eq: theta-vertices}
\end{equation}
so the quartic contact term is independent of $\bar\theta$ while both odd vertices
carry a single factor of $\tan\bar\theta$. The corresponding Feynman rules carry
the combinatorial multiplicities of the functional derivative, namely $4$, $2$ and
$3!$ respectively, so that
\begin{equation}
  V_{aa\pi\pi} = i\,\frac{m_\pi^2}{f_a^2}\,\delta^{bc} ,
  \qquad
  V_{a\pi\pi} = i\tan\bar\theta\,\frac{m_\pi^2}{f_a}\,\delta^{bc} ,
  \qquad
  V_{a^3} = i\tan\bar\theta\,\frac{N_f m_\pi^2 f_\pi^2}{4 f_a^3} .
  \label{eq: vertex-rules}
\end{equation}
Dropping these multiplicities, or using the Lagrangian coefficients themselves as
vertex factors, changes the relative weight of the contact and exchange diagrams
and therefore the sign of their interference. The model of this paper has
$\bar\theta=0$ in both mass cases, so only the contact term survives and every
expression below is to be read at $\tan\bar\theta=0$.

The conversion $\pi\pi\to aa$ receives four contributions, namely the $s$-channel
ALP exchange through the $a^3$ vertex, the $t$- and $u$-channel pion exchange
through two $a\pi\pi$ vertices, and the contact $a^2\pi\pi$ vertex. Adding them
before squaring, and writing the result so that the Bose symmetry of the two
identical ALPs in the final state is manifest,
\begin{equation}
  |\mathcal{M}|^2_{\pi\pi\to aa}
  = \frac{m_\pi^4}{f_a^4}\,\Big|\,\mathcal{D}(s,t,u)\,\Big|^2 ,
  \label{eq: pipi-aa}
\end{equation}
\begin{equation}
  \mathcal{D}(s,t,u)
  = 1 - \tan^2\!\bar\theta
    \left[\frac{N_f m_\pi^2 f_\pi^2}{4 f_a^2\,(s-m_a^2)}
        + \frac{m_\pi^2}{t-m_\pi^2}
        + \frac{m_\pi^2}{u-m_\pi^2}\right] ,
  \label{eq: Dfun}
\end{equation}
where $u=2m_\pi^2+2m_a^2-s-t$ and $\mathcal{D}$ is symmetric under
$t\leftrightarrow u$ by inspection. The two pion-exchange diagrams enter with the
same sign as the contact term, since a tree with two vertices and one propagator
carries $i^3=-i$ against the contact's $+i$ while $t-m_\pi^2$ and $u-m_\pi^2$ are
negative throughout the physical region, so they interfere constructively. The
$s$-channel term does not. Since $s\geq4m_a^2>m_a^2$ everywhere in the physical
region the bracket is positive there, so that contribution enters with the
opposite sign and interferes destructively. It is suppressed by $f_\pi^2/f_a^2$
relative to the exchange terms and does not overturn them, but the three
contributions do not share a common sign.

For the CP-conserving portal of Sec.~\ref{sec: tied} the odd vertices are absent
and Eq.~\eqref{eq: pipi-aa} collapses to a single contact diagram,
\begin{equation}
  |\mathcal{M}|^2_{\pi\pi\to aa}\Big|_{\bar\theta=0} = \frac{m_\pi^4}{f_a^4} ,
  \label{eq: pipi-aa-contact}
\end{equation}
independent of the scattering angle. This is the amplitude that sets the $3\to2$
dominance floor of Sec.~\ref{sec: tied}, and its angular independence is why that
floor can be quoted as a single power law in $f_a$.

The elastic scattering $\pi a\to\pi a$ is not an independent calculation. It shares
the diagram topology, with the $s$-channel ALP exchange becoming a $t$-channel
exchange and the $t$- and $u$-channel pion exchanges becoming $s$- and $u$-channel
exchanges, so its squared amplitude is the $s\leftrightarrow t$ crossing of
Eq.~\eqref{eq: pipi-aa},
\begin{equation}
  |\mathcal{M}|^2_{\pi a\to\pi a}
  = \frac{m_\pi^4}{f_a^4}\,\Big|\,\mathcal{D}(t,s,u)\,\Big|^2 ,
  \label{eq: pia-pia}
\end{equation}
with $u$ as in Eq.~\eqref{eq: Dfun}. Both poles, at $s=m_\pi^2$ and $t=m_a^2$, lie
outside the physical elastic region. At $\bar\theta=0$ this too reduces to
$m_\pi^4/f_a^4$, so the conversion and the elastic channel have the same constant
squared amplitude and differ only through phase space.

Two further amplitudes in this appendix carry the odd vertices and therefore
vanish with them. The $s$-channel piece of $f\bar f\to aa$ in
Eq.~\eqref{eq: ffaa-s} and the first term of $\gamma\gamma\to aa$ in
Eq.~\eqref{eq: gg-aa} both proceed through $a^3$ and so carry
$\tan^2\!\bar\theta$, while Eq.~\eqref{eq: pipi-ff} and the $\pi\pi\to\gamma\gamma$
amplitude above it proceed through a single $a\pi\pi$ vertex and carry
$\tan^2\!\bar\theta$ as well. None of them contributes in this model,
and the energy injection of Sec.~\ref{sec: relocated} runs through the inverse
decay and the $t$- and $u$-channel piece of Eq.~\eqref{eq: ffaa-tu}, neither of
which involves the portal to the pions. The one channel that does reach the pions
without an odd vertex is $f\bar f\to a\pi\pi$, in which an $s$-channel ALP feeds
the contact term of Eq.~\eqref{eq: theta-vertices}. It contributes at the level of
$10^{-3}$ of the injection at the benchmark and is included in the numerics.

\section{Thermal averages and collision integrals}
\label{app: collision}

\subsection*{The thermally averaged cross section}

The thermal average of a $2\to2$ cross section is
\begin{equation}
  \braket{\sigma v}
  = \frac{S}{2T\,K_2(m_1/T)K_2(m_2/T)}
    \int_{s_{\rm min}}^{\infty}\!\! ds\;\sigma(s)\,
    \frac{F^2(m_1,m_2,s)}{m_1^2 m_2^2\sqrt s}\,K_1(\sqrt s/T) ,
  \label{eq: thermal average}
\end{equation}
where $K_{1,2}$ are modified Bessel functions of the second kind and the
remaining quantities are
\begin{align}
  F(m_1,m_2,s) &= \frac{\sqrt{\big(s-(m_1+m_2)^2\big)\big(s-(m_1-m_2)^2\big)}}{2} ,
  \nonumber\\
  s_{\rm min} &= {\rm max}\big[(m_1+m_2)^2,(m_3+m_4)^2\big] ,
\end{align}
with the symmetry factor $S$ equal to $1/2$ for identical initial particles, which
avoids double counting the pair, and to unity otherwise. In the numerical implementation the upper limit of the
integral is set to $T_{\rm RH}^2$, where $T_{\rm RH}$ is the reheating
temperature, which is taken to be the maximum temperature of the SM
bath.

For the small-argument regime that controls the relativistic part of the
injection we use the expansions
\begin{equation}
  K_1(x) \simeq \frac{4 + 2x^2\log\big(xe^{\gamma-1/2}/2\big)}{4x} ,
  \qquad
  K_2(x) \simeq \frac{4-x^2}{2x^2} ,
\end{equation}
valid for $x\ll1$, where $\gamma$ is the Euler--Mascheroni constant. Elsewhere the
Bessel functions are evaluated in exponentially scaled form, which keeps the
integrand within machine range over the whole temperature scan.

\subsection*{The energy-transfer collision integral}

The source term of the temperature equation~\eqref{eq: boltz-T} is the collision
integral
\begin{equation}
  C_E \equiv \int\frac{d^3p}{(2\pi)^3}\,C[f(p,t)]
  = g_1g_2\int\frac{d^3p_1}{(2\pi)^3}\frac{d^3p_2}{(2\pi)^3}\,
    f_1 f_2\, v_{\text{M\o l}}\,\mathcal{E}(\vec p_1,\vec p_2) ,
\end{equation}
where $g_{1,2}$ are the internal degrees of freedom of the incoming particles,
$v_{\text{M\o l}}=F/E_1E_2$ is the M\o ller velocity and
$\mathcal{E}(\vec p_1,\vec p_2)$ is the energy transferred per collision. We take
Maxwell--Boltzmann statistics throughout, so that $f_i=e^{-E_i/T}$.

For a process with non-identical particles in the initial and final states the
energy transfer is
\begin{equation}
  \mathcal{E}(\vec p_1,\vec p_2)
  = \frac{1}{4F}\int\frac{d^3p_3}{(2\pi)^3}\frac{1}{2E_3\,2E_4}\,
    |\mathcal{M}|^2\,(2\pi)\,\delta(E_1+E_2-E_3-E_4)\,\Delta E_{\rm tr} ,
\end{equation}
where $\Delta E_{\rm tr}=E_3$ is the energy carried into the dark sector. The
argument of the delta function is
\begin{equation}
  g(p_3) = E_+ - E_3
    - \big(m_4^2 + E_+ - s + p_3^2 - 2p_1p_3c_{13} - 2p_2p_3c_{23}\big)^{1/2} ,
\end{equation}
where $E_+=E_1+E_2$ and $c_{ij}=\cos\theta_{ij}$. Using the identity
$\delta(g(x))=\sum_i\delta(x-x_i)/|g^\prime(x)|_{x\to x_i}$ and integrating over
$p_3$ leaves a single angular integral,
\begin{equation}
  \mathcal{E}(\vec p_1,\vec p_2)
  = \frac{1}{8\pi F}\int_{-1}^{+1}\! dc_{13}\;
    \frac{p_3^2\,|\mathcal{M}|^2}{4E_4\,|g^\prime(p_3)|_{p_3\to p_3^0}} ,
\end{equation}
where the Jacobian is
\begin{equation}
  g^\prime(p_3)
  = -\frac{p_3}{\sqrt{p_3^2+m_3^2}}
    - \frac{p_3 - p_1c_{13} - p_2c_{23}}
           {\big(m_4^2 + E_+ - s + p_3^2 - 2p_1p_3c_{13} - 2p_2p_3c_{23}\big)^{1/2}} ,
\end{equation}
and $p_3^0$ is the root of $g(p_3)=0$,
\begin{align}
  p_3^0 =\; & -\frac{1}{2\big(E_+^2 - p_1^2c_{13}^2 - 2p_1p_2c_{13}c_{23} - p_2^2c_{23}^2\big)}
    \nonumber\\
  &\times \bigg(E_+^2\Big(-4E_+^2m_3^2 + m_3^4 - 2m_3^2m_4^2
     + 4m_3^2p_1^2c_{13}^2 + 8m_3^2p_1p_2c_{13}c_{23} \nonumber\\
  &\qquad + 4m_3^2p_2^2c_{23}^2 + 2m_3^2 s + m_4^4 - 2m_4^2 s + s^2\Big)^{1/2}
     + p_1c_{13}\big(m_3^2 - m_4^2 + s\big) \nonumber\\
  &\qquad + m_3^2p_2c_{23} - m_4^2p_2c_{23} + p_2 s\,c_{23}\bigg) ,
  \label{eq: p30}
\end{align}
with the two angles related by $\theta_{23}=\theta_{12}+\theta_{13}$, so that
$c_{23}=c_{12}c_{13}-\sin\theta_{12}\sin\theta_{13}$. In the centre-of-mass frame
$\theta_{12}=\pi$ and therefore $c_{23}=-c_{13}$, and the remaining kinematics
reduce to functions of $s$ alone,
\begin{align}
  E_1^* &= \frac{s+m_1^2-m_2^2}{2\sqrt s} , &
  E_2^* &= \frac{s-m_1^2+m_2^2}{2\sqrt s} , \nonumber\\
  E_3^* &= \frac{s+m_3^2-m_4^2}{2\sqrt s} , &
  E_4^* &= \frac{s-m_3^2+m_4^2}{2\sqrt s} ,
\end{align}
with $p_i^*=\sqrt{E_i^{*2}-m_i^2}$ and
$t=m_1^2+m_3^2-2\big(E_1^*E_3^*-p_1^*p_3^*c_{13}^*\big)$.

Following the treatment of the collision terms in
Refs.~\cite{Gondolo:1990dk,Edsjo:1997bg}, the phase-space integral can be written
in the variables $E_\pm=E_1\pm E_2$ and $s$,
\begin{equation}
  C_E = \frac{g_1g_2}{32\pi^4}\int dE_+\,dE_-\,ds\;
        e^{-E_+/T}\,F\,\mathcal{E}(\vec p_1,\vec p_2) .
\end{equation}
Since $\mathcal{E}$ depends only on $s$ in the centre-of-mass frame and the
exponential depends only on $E_+$, the $E_-$ integral can be done at once. The
kinematic limits are
\begin{equation}
  \left|E_- + E_+\left(\frac{m_1^2-m_2^2}{s}\right)\right|
  < \frac{2F}{s}\sqrt{E_+^2-s} ,
\end{equation}
which gives $\int dE_- = 4F\sqrt{E_+^2-s}/s$. The collision integral then reduces
to the two-dimensional form used in the code,
\begin{equation}
  C_E = \frac{g_1g_2}{8\pi^4}
        \int_{(m_1+m_2)^2}^{\infty}\!\! ds
        \int_{\sqrt s}^{\infty}\!\! dE_+\;
        \frac{F^2\sqrt{E_+^2-s}}{s}\,e^{-E_+/T}\,\mathcal{E}(s) .
  \label{eq: CE final}
\end{equation}

\subsection*{The temperature derivative of the energy density}

The dark temperature is not an independent input, so it must be reconstructed
from the energy injected into the hidden sector. While the ALPs and the dark
pions are ultra-relativistic we have $P^\prime=\rho^\prime/3$, and the time
derivative can be traded for a temperature derivative through
$\partial_t\simeq-HT\partial_T$. The energy-balance equation~\eqref{eq: boltz-T}
then becomes
\begin{equation}
  \frac{\partial\rho^\prime}{\partial t} + 4H\rho^\prime
  \simeq -H\left(T\frac{\partial\rho^\prime}{\partial T} - 4\rho^\prime\right)
  = -HT\rho\,\frac{\partial}{\partial T}\left(\frac{\rho^\prime}{\rho}\right)
  = C_E ,
\end{equation}
where the last step uses $\rho\propto T^4$ for the SM bath.
Integrating from the reheating temperature down to $T$ gives
\begin{equation}
  \frac{\rho^\prime}{\rho}(T)
  = \int_T^{T_{\rm RH}}\! \frac{d\tilde T}{H(\tilde T)\,\tilde T\,\rho(\tilde T)}\,
    C_E(\tilde T) ,
\end{equation}
which is Eq.~\eqref{eq: rhop} of the main text. The temperature ratio follows as
$\xi\simeq(\rho^\prime/\rho)^{1/4}$ up to the ratio of effective relativistic
degrees of freedom. Throughout this paper $\xi$ is quoted at $T=3m_\pi$, the
temperature at which the coupled integration begins; below it the dark
temperature is carried by Eq.~\eqref{eq: boltz-T} rather than by this estimate.

\section{Numerical implementation and validation}
\label{app: numerics}

The coupled system $(n_\pi,n_a,T^\prime)$ of
Eqs.~\eqref{eq: boltz-npi}--\eqref{eq: boltz-T} is integrated in the variable
$x=m_\pi/T$ with an implicit stiff solver. Three features of the problem require
care.

The first is the cannibal phase, in which $\Gamma_{3\to2}$ exceeds $H$ by many
orders and the number-density equation becomes stiff. We remove the stiffness by
a change of variables rather than by an approximation, integrating
$\delta_i=\ln(n_i/n_i^{\rm eq}(T^\prime))$ and $\ln(T^\prime/m_\pi)$ against
$\ln z$ with $z=m_\pi/T$. Dividing each collision term by its own density on
paper makes the redshift and equilibrium-tracking pieces cancel symbolically, so
that every source appears as ${\rm expm1}(\delta)$ and vanishes identically at
$\delta=0$. Chemical equilibrium is then the exact fixed point of the system, the
state variables stay of order unity where the densities span sixty orders of
magnitude, and no separation between the injection and the cannibal phase has to
be imposed.

The second is the dynamic range of the thermal averages. The Bessel functions in
Eq.~\eqref{eq: thermal average} are evaluated in exponentially scaled form and the
$s$ integral is performed in $\log s$, which keeps the integrand bounded across the
$T_{\rm RH}$ to $m_\pi$ range.

The third is the reconstruction of $\xi$. The injection
integral~\eqref{eq: CE final} is accumulated from $T_{\rm RH}$ downward and
cross-checked against the closed-form scaling of
Eqs.~\eqref{eq: xi-scaling-open} and \eqref{eq: xi-scaling-closed} by rescaling the
portal coupling and confirming that $\xi$ scales as the expected power. Where this closed form is used in place of the
coupled solve, as in drawing Fig.~\ref{fig: mesa}, the lower limit of the
integration matters, because the integrand is sharply peaked. With the inverse
decay open it behaves as $T^{-9/2}e^{-m_a/T}$ and peaks at $T=2m_a/9$, well below
the dark pion mass, so truncating at a temperature of order $m_\pi$ discards most
of the injected energy; we integrate a further factor of twenty below the peak,
by which point the accumulated integral has converged. The coupled solve carries
no such truncation: it sets the initial condition at $T=3m_\pi$ from the energy
injected above that temperature and then continues to inject through the source
terms of Eqs.~\eqref{eq: boltz-npi}--\eqref{eq: boltz-T}. The effective
relativistic degrees of freedom entering $\rho$ and $s$ are taken from the
tabulation of Ref.~\cite{Saikawa:2018rcs} throughout rather than held fixed at a
representative value, since the surviving integrand sits near the QCD crossover
where $g_*$ is changing rapidly.

We validate the solver in two limits. Setting $T^\prime=T$ by hand reproduces the
standard SIMP relic curve of Ref.~\cite{Hochberg:2014kqa}, with the perturbativity
edge $m_\pi/f_\pi=2\pi$ reached near $m_\pi\simeq400\MeV$ for the group-theory
factors adopted in Sec.~\ref{sec: model}. Switching off the $3\to2$ term recovers
the pure freeze-in yield $\Omega_\pi h^2\propto g_{aff}^2$ quoted in
Sec.~\ref{sec: regimes}. All runs take $N_c=2$ and the Froggatt--Nielsen charges of
Table~\ref{tab: charges}. The benchmark of Sec.~\ref{sec: tied} is at
$m_\pi=140\MeV$ with $m_a=1.25\,m_\pi$. Figures~\ref{fig: wzw rates} and
\ref{fig: regimemap} are drawn at $m_\pi=100\MeV$ instead, as generic
illustrations of the rate hierarchy and the regime boundaries; both depend on
$m_\pi$ only through the overall scale of the rates, and neither is used to fix a
number quoted in the text.